\documentclass[11pt]{article} 
\usepackage{amsmath,amsfonts,amssymb,epsfig,wick,a4}
\newcommand{\rightslash}{\! \stackrel{\rightarrow}{\!\slash{\partial}}}
\newcommand{\leftslash}{\! \stackrel{\leftarrow}{\!\slash{\partial}}}
\newcommand{\leftDslash}{\! \stackrel{\leftarrow}{\slash{\! D}}}

\newcommand{\kslash}{\slash{k}}
\begin{document}

\def\be{\begin{equation}}
\def\ee{\end{equation}}
\def\bea{\begin{eqnarray}}
\def\eea{\end{eqnarray}}
\def\rarr{\rightarrow}
\def\C{{\rm\kern.24em
    \vrule width.02em height1.4ex depth-.05ex
    \kern-.26em C}}
\def\N{{\rm I\kern-.18em N}}
\def\R{{\rm I\kern-.21em R}}
\def\Z{{\rm\kern.26em
    \vrule width.02em height0.5ex depth 0ex
    \kern.04em
    \vrule width.02em height1.47ex depth-1ex
    \kern-.34em Z}}
\def\d{{\rm\kern.22em
    \vrule width.02em height1.0ex depth0ex
    \kern-.24em d}}
\def\nn{\nonumber}
\def\fr{\frac}
\renewcommand\slash[1]{\not \! #1}
\newcommand\qs{\!\not \! q}
\def\del{\partial}
\def\gam{\gamma}
\newcommand\vphi{\varphi}
\def\tr{\mbox{tr}\,}
\newcommand\qt{\tilde{q}}
\newcommand\ns{\!\not \! n}
\newcommand\lto{\longrightarrow}
\newcommand\real{\mbox{Re}\,}
\newcommand\imag{\mbox{Im}\,}
\newcommand\nin{\noindent}
\newcommand\lbr{\left(}
\newcommand\rbr{\right)}
\newcommand\lbk{\left[}
\newcommand\rbk{\right]}
\newcommand\lbc{\left\{}
\newcommand\rbc{\right\}}
\newcommand\bmat{\boldmath}
\newcommand\ubmat{\unboldmath}
\newcommand\mb{\mbox}
\newcommand{\vDelta}{\vec\Delta}
\newcommand{\vdelta}{\vec\delta}

\def\rarr{\rightarrow}
\def\del{\partial}
\def\pbar{\bar{p}}
\def\zbar{\bar{z}}
\def\rhobar{\bar{\rho}}
\def\kf{{\bf k}}
\def\qf{{\bf q}}
\def\lf{{\bf l}}
\def\vf{{\bf v}}
\def\wf{{\bf w}}
\def\Af{\mbox{\bf A}}  
\def\Vf{\mbox{\bf V}}  
\def\Ff{\mbox{\bf F}}  
\def\ca{{\cal A}}
\def\cA{{\cal A}}
\def\cb{{\cal B}}
\def\ccal{{\cal C}}
\def\dcal{{\cal D}}
\def\cD{{\cal D}}
\def\cS{{\cal S}}
\def\cO{{\cal O}}
\def\cP{{\cal P}}
\def\cx{{\cal X}}
\def\cy{{\cal Y}}
\def\cz{{\cal Z}}
\def\rhobar{\bar{\rho}}
\def\C{{\rm\kern.24em
    \vrule width.02em height1.4ex depth-.05ex
    \kern-.26em C}}
\def\N{{\rm I\kern-.18em N}}
\def\P{{\rm I\kern-.25em P}}
\def\R{{\rm I\kern-.21em R}}
\def\Z{{\rm\kern.26em
    \vrule width.02em height0.5ex depth 0ex
    \kern.04em
    \vrule width.02em height1.47ex depth-1ex
    \kern-.34em Z}}
\def\nn{\nonumber}
\def\fr{\frac}
\renewcommand\slash[1]{\not \! #1}


\begin{titlepage}
\begin{flushright}
HD-THEP-04-17\\
IFUM-789-FT\\
hep-ph/0404254
\\
\end{flushright}
\vfill
\begin{center}
\boldmath
{\LARGE{\bf Towards a Nonperturbative Foundation}}\\[.2cm]
{\LARGE{\bf of the Dipole Picture:}}\\[.2cm]
{\LARGE{\bf I. Functional Methods}}
\unboldmath
\end{center}
\vspace{1.2cm}
\begin{center}
{\bf \Large
Carlo Ewerz\,$^{a,b,1}$, Otto Nachtmann\,$^{a,2}$
}
\end{center}
\vspace{.2cm}
\begin{center}
$^a$
{\sl
Institut f\"ur Theoretische Physik, Universit\"at Heidelberg\\
Philosophenweg 16, D-69120 Heidelberg, Germany}
\\[.5cm]
$^b$
{\sl
Dipartimento di Fisica, Universit{\`a} di Milano and INFN, Sezione di Milano\\
Via Celoria 16, I-20133 Milano, Italy}
\end{center}                                                                 
\vfill
\begin{abstract}
\noindent
This is the first of two papers in which we study real and 
virtual photon-proton scattering in a nonperturbative framework. 
We classify different contributions to this process and identify 
the leading contributions at high energies. 
We then study the renormalisation of the photon-quark-antiquark 
vertex that occurs in the leading contributions. 
We find something like the dipole picture in one of these 
contributions but also find two correction terms which can potentially 
become large at small photon virtualities. 
In the second paper we will discuss the additional 
approximations and assumptions that are necessary to obtain the 
dipole model of high energy scattering from the results found here. 
\vfill
\end{abstract}
\vspace{5em}
\hrule width 5.cm
\vspace*{.5em}
{\small \noindent 
$^1$ email: C.Ewerz@thphys.uni-heidelberg.de \\
$^2$ email: O.Nachtmann@thphys.uni-heidelberg.de 
}
\end{titlepage}

\section{Introduction}
\label{sec:intro}

The dipole picture of deep inelastic scattering at high energies 
\cite{Nikolaev:1990ja}--\cite{Mueller:gb} 
is a popular and successful way of describing 
and analysing the HERA data on electron-proton 
scattering, that is in particular quasi-real and virtual photon-proton 
scattering. 
In the recent past it has turned out 
that the framework of the dipole picture is well 
suited for studying the region of relatively small photon virtualities 
$Q^2$, that is the interesting transition region from perturbative 
to nonperturbative QCD. Particular interest 
concentrates on the problem of finding possible saturation effects 
occurring in that transition region. In essence, the dipole picture 
describes the deep inelastic photon-proton interaction as a 
two-step process in which the photon first splits into a quark-antiquark 
pair -- a colour dipole -- 
which then in the second step scatters off the proton. 
The first foundations on which the dipole picture rests 
were already laid in \cite{Gribov:1968gs,Ioffe:1969kf}.
For a general discussion of the dipole picture 
and for further references see \cite{Donnachie:en}. 
One key problem in applying the dipole model in the transition 
region is to find a suitable description of the second step of the 
reaction, in other words to find the correct behaviour of the 
dipole-proton cross section. Models for this cross section 
have been developed and tested in inclusive as well as 
diffractive scattering processes, see for example 
\cite{Golec-Biernat:1998js}--\cite{Golec-Biernat:2006ba}. 
Particularly popular is the Golec-Biernat--W\"usthoff model 
\cite{Golec-Biernat:1998js}--\cite{Golec-Biernat:2001mm} 
which already in its first and simplest version gave  
a surprisingly good description of the HERA data. 
An important improvement of that model was obtained 
in \cite{Bartels:2002cj} due to the inclusion of the correct 
logarithmic scaling violations according to DGLAP evolution 
at large $Q^2$ while preserving the success of the model at 
low $Q^2$. More recently, also the problem of describing 
the dependence of the dipole cross section on the impact 
parameter and hence the question of the $t$-distribution 
of the cross section has been addressed 
\cite{Bartels:2003yj,Kowalski:2003hm}. 

The two-step nature of the 
scattering process described above is most easily visualised 
in a perturbative situation, and much of the theoretical development 
of the dipole model has been guided by the perturbative picture 
of deep inelastic scattering in QCD at high energies. 
Loosely speaking the dipole model tries to approach the 
transition to the nonperturbative region of small momenta 
from the perturbative side. Although its phenomenological 
success is rather encouraging it is not a priori clear that this 
method is in fact admissible. It is therefore very important 
to study whether there can be contributions 
to the cross section at low $Q^2$ which are not compatible 
with the formalism of the dipole model, that is contributions 
which cannot be correctly accommodated despite the freedom 
in modelling the low-$Q^2$ dipole-proton cross section. 
It is one aim of the present paper and the companion paper 
\cite{Ewerz:2006vd} (hereafter referred to as II) to address this 
question. One of our main results will in fact be the identification of 
two contributions to the amplitude which are not contained 
in the simple dipole picture. Both of these correction terms 
are small at large $Q^2$ but can become large at small $Q^2$. 
A second question which we consider to be important and which 
should be addressed in a nonperturbative framework is the following. 
It is frequently asserted that some sort of saturation must occur 
in real or virtual photon-proton scattering at high energies 
due to the Froissart-Martin-Lukaszuk bound 
on total cross sections. However, this bound applies only to 
hadron-hadron scattering and a priori is not valid for a 
current-induced reaction (see for instance \cite{Donnachie:en}). 
It is usually argued that the dipole formula allows one 
to consider photon-proton scattering in essence as a 
hadron-proton reaction. But if one really wants to substantiate 
this type of reasoning and maybe some day derive an analogue 
of the Froissart-Martin-Lukaszuk bound for $\gamma p$ 
scattering one certainly should have a solid foundation for 
something like the dipole picture. 
It is our opinion that in order to achieve this goal one should 
not rely on perturbation theory alone. 

In this work we start from a description of the real or virtual 
photon-proton scattering 
process in a generically nonperturbative framework 
developed by one of us in \cite{Nachtmann:ua,Nachtmann:1996kt}. 
This framework allows us to find a classification of different 
contributions to the cross section and to identify the 
approximations and assumptions that are necessary to 
arrive at the dipole model. We are able to identify two 
corrections to the dipole picture which can become 
large especially at low photon virtualities. 
In addition we find that our 
approach is also well suited to study two important problems 
which in our opinion have not yet been sufficiently 
appreciated so far. The first problem is that in QCD 
a mass-shell condition for quarks does not exist due to 
confinement. This is clearly not expected to be a serious 
problem in perturbative situations. However, if one wants 
to study high energy scattering in the nonperturbative 
domain this problem has to be taken very seriously. 
We will not be able to solve this deep problem 
in the present papers, but we hope to provide a framework 
in which at least the effects of assuming a mass-shell condition 
for quarks can be studied in more detail. 
The other important issue to which we would like to draw 
some attention concerns the electromagnetic gauge invariance 
in photon-proton scattering. It is well-known that gauge invariance 
in general imposes very strong restrictions on the amplitudes 
and on their factorisation. 
This has for example been discussed for the case of diffractive 
$\rho$ production at high energies in \cite{Hebecker:1997rv}. 
In II we find that important corrections 
to the simple dipole model are in fact required in order to 
ensure gauge invariance. It turns out that only at asymptotically high 
energies the simple dipole formula is obtained from the nonperturbative 
framework as a gauge invariant expression. This immediately 
raises the question whether presently accessible energies 
are large enough for the dipole picture to give a 
gauge invariant description in a sufficiently good approximation. 
Our framework should make it possible to address this important 
question in more detail. 

The two papers are organised as follows. In the first paper 
we describe the general nonperturbative framework of our 
considerations and derive a classification of the contributions 
to the process of real or virtual Compton scattering on a 
proton. After discussing the relative importance of the 
different contributions we proceed to study in detail those 
which we expect to be leading in the high energy limit. 
In the further discussion we then focus on these contributions 
and discuss in particular the renormalisation of the 
photon-quark-antiquark vertex. 
The high energy limit of the amplitude obtained so far is then 
studied in the second paper. There we discuss 
how the usual formula for photon-proton scattering in 
the dipole picture emerges. Also in the second paper we 
discuss the important issue of gauge invariance. 
In particular we study how gauge invariance is manifest 
in the leading contribution to the scattering process. 
We then show that in the high energy limit different 
terms in the amplitude become gauge invariant separately. 
The perturbative wave function of the photon is derived and 
the consequences of gauge invariance for its correct definition  
are pointed out. Finally, we discuss some properties of the 
wave function of the photon and derive some 
simple phenomenological consequences of the dipole picture 
which can help to determine the limits of its applicability. 
We present our conclusions in II and point out some 
questions that in our opinion deserve further study. 
Some technical details of our calculations are presented 
in the appendices of the two papers. 

\section{Nonperturbative techniques and the dipole picture}
\label{nonpertsect}

Here we analyse photon induced reactions using nonperturbative 
techniques introduced in \cite{Nachtmann:ua,Nachtmann:1996kt}. 
The aim is to see if one can 
justify the dipole picture for high energies or if modifications of the dipole 
picture are necessary. Our method is quite general and can be applied to 
many processes. 
\vspace*{.3cm}
\begin{figure}[hb]
\begin{center}
\includegraphics[width=6.5cm]{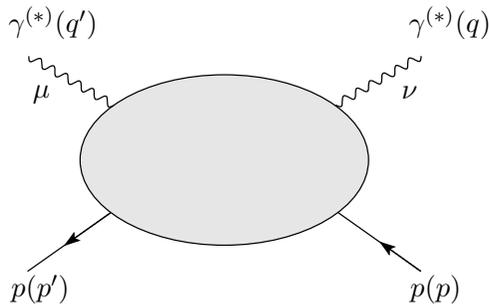}
\caption{Compton scattering on a proton
\label{fig1}}
\end{center}
\end{figure}
For definiteness we consider real or virtual Compton 
scattering on a  proton (see figure \ref{fig1}):
\begin{equation}
\label{2.1}
\gamma^{(*)}(q)+p(p)\,\,\,\longrightarrow\,\,\,
\gamma^{(*)}(q^\prime)+p(p^{\prime}) \,,
\end{equation}
where we indicate the four-momenta in brackets. 
For $q'=q$ the absorptive part of the Compton amplitude can 
be related to spin-dependent and spin-independent structure functions. 
In the present papers we always work in leading 
order of the expansion in powers of 
the electromagnetic coupling $\alpha_{\rm em}$. 
Note that throughout this paper the figures are drawn such that 
the incoming particles are on the right hand side. We have 
chosen this ordering in order to make the correspondence 
of the figures with the equations as close as possible -- 
despite the fact that the opposite choice is probably more 
common in the literature. 
We suppose
\begin{equation}
q^2=-Q^2\leq 0 \,,\:\:\:\:\:
q^{\prime\,2}=-Q^{\prime\,2}\leq 0 \,,
\end{equation}
but do not require $Q^2=Q^{\prime\,2}$. The usual Mandelstam variables are
\begin{eqnarray} 
\label{2.2a}
s&=&(p+q)^2=(p^{\prime}+q^{\prime})^2 \,,\nonumber\\
t&=&(p-p^{\prime})^2=(q^{\prime}-q)^2 \,,\\
u&=&(p-q^{\prime})^2=(p^{\prime}-q)^2 \,.
\nonumber
\end{eqnarray}
Our procedure is rather general 
and allows us to treat the Compton amplitude for the case of 
real and virtual photons on the same footing. In the case of 
virtual photons our calculations apply 
to transversely as well as to longitudinally polarised photons. 
However, for the latter there are some subtleties in the high 
energy limit, as we will explain in II. 
We will at first keep to the cartesian indices $\mu,\nu=0,1,2,3$ 
for the photon polarisations. 

Before going into the technical details let us now give 
a short non-technical description of our method.
We use the reduction formula to represent the amplitude for our process 
(\ref{2.1}) as an integral over vacuum expectation values of the currents 
and the fields for the protons. Then these vacuum expectation values are 
expressed as functional integrals in the usual way. Since the Lagrangian 
of QCD is bilinear in the quark degrees of freedom, $q$ and $\bar q$, 
we can perform the integrations over $q$ and $\bar q$. This allows us 
to give a classification of contributions to our amplitude in diagrammatic 
language according to the topology of the quark line skeleton, see 
figure \ref{fig2} below.  
There the quark lines correspond to full quark propagators in a given 
external gluon field and the shaded blobs indicate the functional 
integration over all external gluon fields with a measure given below. 
The usual expectation is that at high energies the contribution shown 
in figure \ref{fig2}a will dominate. 
We show that also the diagram of figure \ref{fig2}b can give a leading 
contribution at high energies. 
We study first figure \ref{fig2}a in detail. 
We `cut' the quark lines after the vertex where the photon 
$\gamma(q,\nu)$ enters the diagram and insert suitable factors 1, 
given by the Dirac operator times the free Green's functions of the 
quarks. Integration by parts and the representation of the free Green's 
functions of the quarks as dyadic products of quark and antiquark 
wave functions lead to four terms one of which can be interpreted as 
expected in the dipole picture. The photon $\gamma (q,\nu)$ splits into 
a quark-antiquark pair, which then interacts in all possible ways with 
the proton, see figures \ref{fig3}a and \ref{fig4}a below. 
However, these quarks and antiquarks are still off their energy shell, 
and one has to integrate over all these off-shell energies. 
But at high energies the leading 
contribution comes from the on-shell quarks. This corresponds to the 
contribution of a certain pinch singularity in the amplitudes 
considered as functions in the (off-shell) energy plane. 
Adding now the contribution of the diagram of figure \ref{fig2}b, 
treated in a similar way, 
we have indeed described the original amplitude for the process (\ref{2.1}) 
at high energies as a two step process. The  leading term 
of the amplitude is given as a convolution of the bare $\gamma q \bar q$ 
vertex function with a scattering amplitude for a bare $q \bar q$ pair 
on the proton. It remains to introduce the renormalised quantities in 
place of the bare ones. This we do by using the Dyson-Schwinger equation 
for the $\gamma q \bar q$ vertex function which contains all higher 
order QCD corrections to the vertex. Here our approach is inspired 
by the standard treatment of overlapping divergences in QED. 
Finally, we arrive at the expression shown graphically in 
figures \ref{fig6} and \ref{fig9} below. 
The leading contribution to  
the amplitude for the reaction (\ref{2.1}) at high energies is given 
by two terms. The first one is a convolution of the renormalised 
$\gamma q\bar q$ vertex function, describing the splitting of the photon 
to the $q\bar q$ pair, with the renormalised amplitude for 
$q\bar q +p\rightarrow \gamma(q^\prime,\mu)+p$. The second 
term describes rescattering corrections. 
Thus, on the side of the colour dipole amplitude, the contribution 
${\cal T}^{(b)}$ from the diagram of figure \ref{fig2}b must be added, 
and on the side of the incoming photon, we find the rescattering 
correction. These additional terms are also leading at high 
energies and could be important for phenomenology especially at 
small photon virtualities. 

In addition to deriving this general structure of the amplitude 
for (\ref{2.1}) at high energies we discuss in II the problem 
of electromagnetic gauge invariance and the related subtleties for 
the amplitudes of longitudinally polarised virtual photons. 
In summary, from the contribution of figure \ref{fig2}a and \ref{fig2}b 
we find something like the dipole picture at high energies. 
But usually only the term ${\cal T}^{(a)}$ from the diagram 
of figure \ref{fig2}a times the photon vertex function is discussed 
in the dipole model. We find important additional terms as mentioned above. 

A key step in our procedure is to cut the amplitude next to the vertex 
of the incoming photon and to insert a factor 1 consisting of the 
Dirac operator and of Green's functions of the quarks. 
A similar step has of course been used more or less explicitly 
in many studies of the Compton amplitude or of diffractive 
photon-proton scattering. Many of those investigations are, 
however, restricted to the case of virtual photons in which 
perturbation theory can be applied. In our method this step 
is used in the framework of a nonperturbative description of the 
real and the virtual Compton amplitude. 
But several technical aspects of our method clearly 
resemble those studies, see for example 
\cite{Buchmuller:1996xw,Hebecker:1999ej} and also 
\cite{McLerran:1998nk,Venugopalan:1999wu}. 

After this outline of our procedure we now turn to the actual 
calculations. These involve sometimes lengthy formulae. But we hope 
that the figures accompanying them will, nevertheless, make them 
transparent. We should point out again that in order to make the 
correspondence between the equations and the figures as easy 
as possible all figures are drawn such that they are to be read 
from right to left. 

\subsection{Classification of contributions to real or 
virtual Compton scattering}
The matrix element for the reaction (\ref{2.1}) is
\be
\label{2.3}
\mathcal{M}^{\mu\nu}_{s^{\prime}s}(p^{\prime},p,q)
=\frac{i}{2\pi m_p}\int d^4x~e^{-iqx}
\langle p(p^{\prime},s^{\prime})|\mbox{T}^*J^{\mu}(0)J^{\nu}(x)|p(p,s)\rangle
\,,
\ee
where $m_p$ is the proton mass, $s,s^{\prime}=\pm 1/2$ are the 
spin indices, and $\mbox{T}^*$ is the covariantised T-product.
Our proton states are normalised to
\begin{equation}\label{2.3a}
\langle p(p^{\prime},s^{\prime})|p(p,s)\rangle=2p^0(2\pi)^3
\delta_{s^{\prime}s} \,\delta^{(3)}(\mathbf{p}^{\prime}-\mathbf{p}) \,. 
\end{equation}
All our conventions regarding kinematics, Dirac spinors etc.\ follow \cite{4}.

The hadronic part of the electromagnetic current is denoted by 
\begin{equation}
J^{\lambda}(x)=\sum_q \bar{q}(x)Q_q\gamma^{\lambda}q(x) \,,
\end{equation}
where $q=u,d,\dots$ are the quark field operators and $Q_u=2/3,Q_d=-1/3, \dots$ 
are the quark charges in units of the proton charge.

Let $\psi_p(x)$ be a suitable interpolating field operator for the proton. We can take
$\psi_p$ to contain three quark fields, 
\be
\label{2.5}
\psi_p(x)=\Gamma_{\alpha\beta\gamma}
u_{\alpha}(x)u_{\beta}(x)d_{\gamma}(x) \,,
\ee
and accordingly 
\be
\bar{\psi}_p(x)=\bar{\Gamma}_{\alpha\beta\gamma}
\bar{d}_{\gamma}(x)\bar{u}_{\beta}(x)\bar{u}_{\alpha}(x)\,,
\ee
where $\Gamma_{\alpha\beta\gamma}$ and 
$\bar{\Gamma}_{\alpha\beta\gamma}$ are 
coefficient matrices and 
$\alpha,\beta,\gamma$ summarize Dirac and colour indices. 
Explicit constructions of field operators $\psi_p(x)$ of the type (\ref{2.5}) 
can be found in \cite{Chung:wm,Chung:cc,Ioffe:kw}. Let $Z_p$ be the
proton's wave function renormalisation constant defined by
\begin{equation}\label{2.5a}
\left\langle0|\psi_p(x)|p(p,s)\right\rangle
=\sqrt{Z_p} \, e^{-ipx}u_s(p)\,.
\end{equation}

Using now the LSZ reduction formula \cite{Lehmann:1954rq} and the 
representation of Green's functions by the functional integral we get 
\begin{eqnarray}
\label{2.6}
\mathcal{M}^{\mu\nu}_{s^{\prime}s}(p^{\prime},p,q)&=&
-\frac{i}{2\pi m_pZ_p}
\int d^4y^{\prime}~d^4y~d^4x \, e^{ip^{\prime}y^{\prime}}
\bar{u}_{s^{\prime}}(p^{\prime})
(-i\rightslash_{y'}+m_p)
\\
&&{}
\left\langle\psi_p(y^{\prime})J^{\mu}(0)
e^{-iqx}J^{\nu}(x)\overline{\psi}_p(y)\right\rangle_{G,q,\bar{q}} \,
(i\leftslash_y+m_p)u_s(p)e^{-ipy}
\,.
\nonumber
\end{eqnarray}
Here we use a shorthand notation for the functional integral. For any functional
$F(G,q,\bar{q})$ of gluon and quark fields we write 
\begin{equation}\label{2.7}
\left\langle F(G,q,\bar{q})\right\rangle_{G,q,\bar{q}}=\frac{1}{\mathcal{Z}}\int
\mathcal{D}(G,q,\bar{q})F(G,q,\bar{q})\exp
\left[ i\int d^4x \,{\cal L}_{\mbox{\scriptsize QCD}}(x)\right] \,,
\end{equation}
with 
\begin{equation}\label{2.7a}
\mathcal{Z}=\int \mathcal{D}(G,q,\bar{q})
\exp\left[ i\int d^4x \, {\cal L}_{\mbox{\scriptsize QCD}}(x)\right] \,.
\end{equation}
Of course, on the r.h.s.\ of (\ref{2.6}), the quark-field operators in 
$J^{\mu}$ and $\psi_p$ are to be replaced by the corresponding 
Grassmann integration variables of the functional integral. 
Gauge fixing and Faddeev-Popov terms in the functional integral 
are implied and not written out explicitly.

In (\ref{2.6}) we have used translational invariance to shift the 
argument of the current $J^\mu$ to zero. In the following this 
makes our procedure look somewhat asymmetric 
with respect to the treatment of the two currents $J^\mu$ and 
$J^\nu$. We could equally well make the treatment completely 
symmetric with respect to $J^\mu$ and $J^\nu$ by shifting 
the argument of $\psi_p$ or $\bar{\psi}_p$ to zero in (\ref{2.6}), 
see appendix \ref{appA}. 

Since the Lagrangian of QCD is bilinear in the quark fields, we can 
immediately perform the $q,\bar{q}$ functional integral. 
This leads to a number of terms 
representing all possible contractions of quark and antiquark field 
operators, as is explained in appendix \ref{appA}. 
The resulting terms can be classified in terms of the 
quark line skeleton in the graphical representation of the amplitude. 
The different types of contributions to the amplitude 
are shown pictorially in figure \ref{fig2}, 
the explicit formulae for all terms are given explicitly in 
appendix \ref{appA}. 
\begin{figure}
\begin{center}
\includegraphics[width=14.45cm]{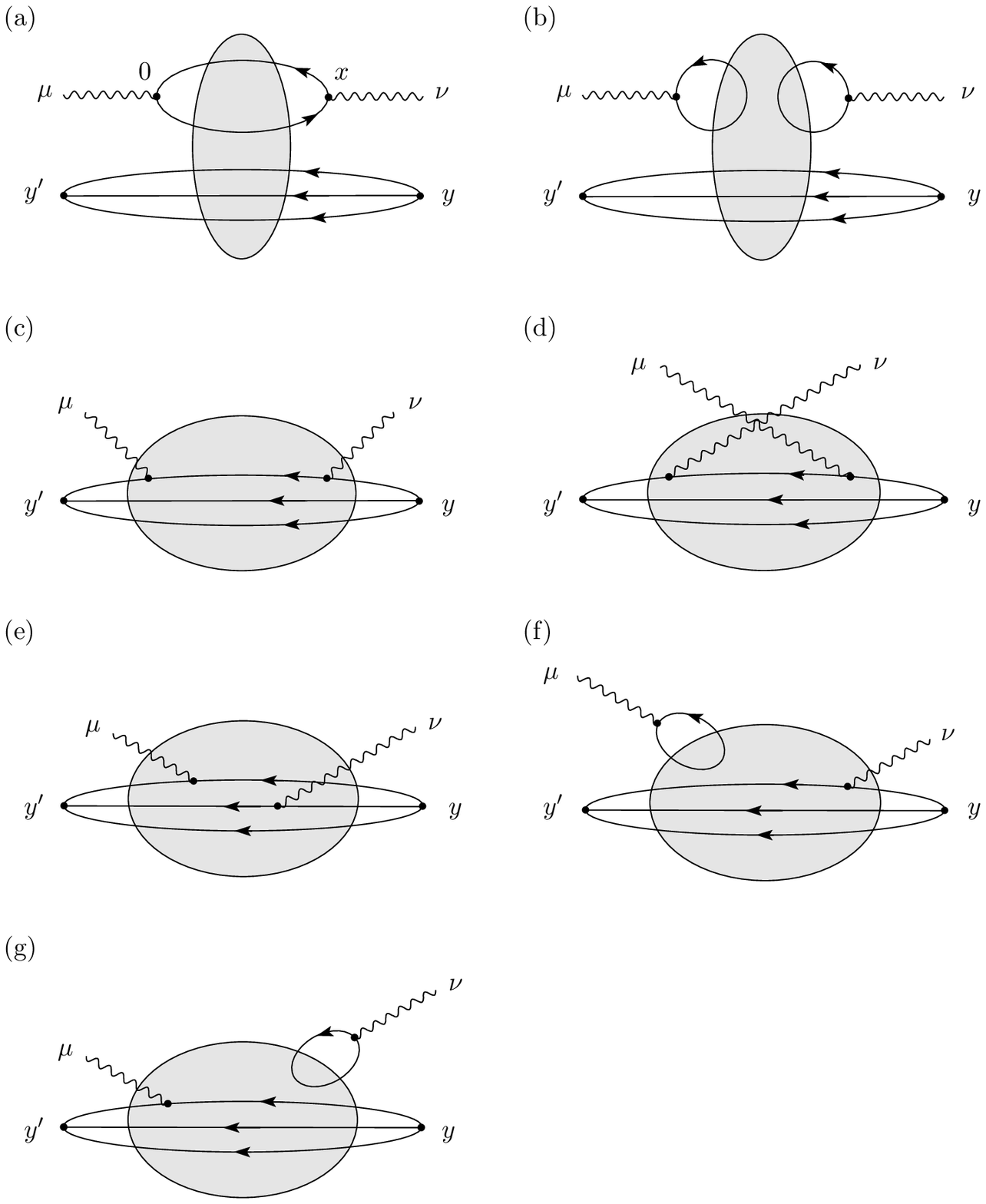}
\vspace*{.3cm}
\caption{The diagrams resulting from the functional integral (\ref{A.7}) 
corresponding to ${\cal J}^{(a)},\dots,{\cal J}^{(g)}$ and to the amplitudes 
${\cal M}^{(a)}, \dots, {\cal M}^{(g)}$ of (\ref{2.7a1})
\label{fig2}}
\end{center}
\end{figure}
In figure \ref{fig2} each
quark line corresponds to a full quark propagator in a given external gluon potential.
An integration over all gluon potentials with a functional integral measure including
the fermion determinant is implied. Corresponding to the diagrams 
of figure \ref{fig2} we get the
following decomposition of $\mathcal{M}^{\mu\nu}_{s^{\prime}s}$ 
(\ref{2.6}): 
\begin{equation}\label{2.7a1}
\mathcal{M}^{\mu\nu}_{s^{\prime}s}(p^{\prime},p,q)=
\mathcal{M}^{(a)\mu\nu}_{s^{\prime}s}(p^{\prime},p,q)
+\ldots+\mathcal{M}^{(g)\mu\nu}_{s^{\prime}s}(p^{\prime},p,q)\,.
\end{equation}
Putting all factors together we get for instance for 
$\mathcal{M}^{(a)\mu\nu}_{s^{\prime}s}$ (see figure \ref{fig2}a)
\begin{eqnarray}\label{2.8}
\mathcal{M}^{(a)\mu\nu}_{s^{\prime}s}(p^{\prime},p,q)&=&
- \frac{i}{2\pi m_pZ_p}\int d^4y^{\prime}~d^4y \,
e^{ip^{\prime}y^{\prime}}\bar{u}_{s^{\prime}}(p^{\prime})(-i\rightslash_{y^{\prime}}
+m_p) 
\\
&&{}
\sum_q Q^2_q  
\Big\langle
\wick{2}{<1\psi_p(y^{\prime})>1{\overline{\psi}}_p(y)}
A^{(q)\mu\nu}(q)\Big\rangle_G \,
(i\leftslash_y+m_p)u_s(p)e^{-ipy} \,,
\nonumber
\end{eqnarray}
where the contraction 
$\wick{2}{<1\psi_p(y^{\prime})>1{\overline{\psi}}_p(y)}$ 
is defined in (\ref{A.10}) and
\be
\label{2.8a}
A^{(q)\mu\nu}(q)=\int d^4x \,
\mathrm{Tr}\left[\gamma^{\mu}S^{(q)}_F(0,x;G)
e^{-iqx}\gamma^{\nu}S^{(q)}_F(x,0;G) \right] 
\,.
\ee
Here $S^{(q)}_F(x,y;G)$ is the propagator for quark flavor $q$ in the 
external gluon potential $G$, 
\begin{equation}\label{A.4}
(i\gamma^{\mu}D_{\mu}-m^{(0)}_q)S^{(q)}_F(x,y;G)=-\delta^{(4)}(x-y)\,.
\end{equation}
In (\ref{2.8}) we denote by 
$\langle \cdot \rangle_G$ the functional integral over the gluonic degrees 
of freedom, including the fermion determinant. That is,
we define for any functional $F[G]$ 
\begin{eqnarray}\label{2.9}
\left\langle F[G]\right\rangle_G&=&\frac{1}{\mathcal{Z}^{\prime}}
\int \mathcal{D}(G)F[G] \,
\prod\limits_q \det\left[-i( i \gamma^{\lambda}
D_{\lambda}-m^{(0)}_q+i\epsilon)\right]\nonumber\\
&&{}
\exp\left[-i\int d^4x \, \frac{1}{2}\mathrm{Tr} 
\left( G_{\lambda\rho}(x)G^{\lambda\rho}(x)\right)
\right]\,,
\end{eqnarray}
where $\mathcal{Z}^{\prime}$ is the normalisation factor obtained from 
the condition
\begin{equation}\label{2.10}
\langle 1 \rangle_G=1 \,.
\end{equation}
Again gauge fixing and Faddeev-Popov terms are implied. 

In a similar way we get from (\ref{A.7a}) and (\ref{A.11}) for 
the amplitude $\mathcal{M}^{(b)\mu\nu}_{s^{\prime}s}$ 
corresponding to figure \ref{fig2}b 
\begin{eqnarray}
\label{Mbexpl}
\mathcal{M}^{(b)\mu\nu}_{s^{\prime}s}(p^{\prime},p,q) &=&
- \frac{i}{2\pi m_pZ_p}\int d^4y^{\prime}~d^4y \,
e^{ip^{\prime}y^{\prime}}\bar{u}_{s^{\prime}}(p^{\prime})(-i\rightslash_{y^{\prime}}
+m_p) 
\\
&&{}
\sum_{q',q} Q_{q'} Q_q  
\Big\langle
\wick{2}{<1\psi_p(y^{\prime})>1{\overline{\psi}}_p(y)} (-1)
\mathrm{Tr} \left[ \gamma^\mu S_F^{(q')} (0,0;G) \right]
\nn \\
&&{} 
\int d^4x \, \mathrm{Tr} \left[ e^{-iqx} \gamma^\nu 
S_F^{(q)} (x,x;G) \right] \Big\rangle_G 
(i\leftslash_y+m_p)u_s(p)e^{-ipy} 
\,.
\nn
\end{eqnarray}
The corresponding expressions for the other contributions 
$\mathcal{M}^{(c)}$ - $\mathcal{M}^{(g)}$ are given 
explicitly in appendix \ref{appA}. 

The methods used here for the classification of different 
contributions to the amplitude can also be applied to other 
current induced reactions at high energies. 
In the case of exclusive reactions for example one 
chooses appropriate interpolating field operators for the 
produced hadrons and can proceed exactly as discussed here 
for the Compton amplitude. 

\subsection{Relative size of different contributions}
\label{sec:relsize}

With the diagrams of figure \ref{fig2} we have given a classification of 
all contributions to the Compton amplitude, see (\ref{2.7a1}). 
At high energies $\sqrt{s}$ we expect the diagrams of figures \ref{fig2}a 
and \ref{fig2}b, that is $\mathcal{M}^{(a)}$ of (\ref{2.8}) and 
$\mathcal{M}^{(b)}$ of (\ref{Mbexpl}), 
to give the dominant contribution. 
Indeed, let us consider the diagrams of figures \ref{fig2}a-\ref{fig2}g 
for large $\sqrt{s}$. In figure \ref{fig2}c for instance we find at Born 
level that a large momentum of order $\sqrt{s}$ has to flow through 
the upper quark line. This leads to a suppression of order $1/\sqrt{s}$ 
relative to the diagrams of figures \ref{fig2}a, \ref{fig2}b. 
Similar arguments apply to the diagrams of figures \ref{fig2}d-\ref{fig2}g. 

\begin{figure}[ht]
\begin{center}
\includegraphics[width=6.1cm]{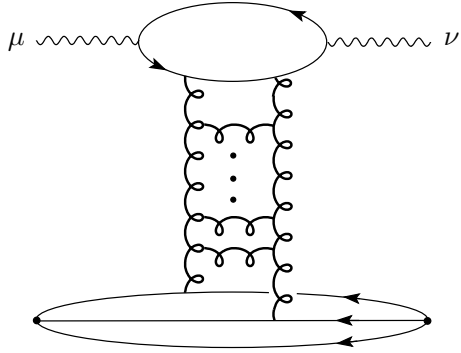}
\vspace*{.3cm}
\caption{Perturbative gluon ladder diagram which is contained in the 
generic diagram of figure \ref{fig2}a for the amplitude $\mathcal{M}^{(a)}$ 
\label{fig200}}
\end{center}
\end{figure}

\begin{figure}[ht]
\begin{center}
\includegraphics[width=7.1cm]{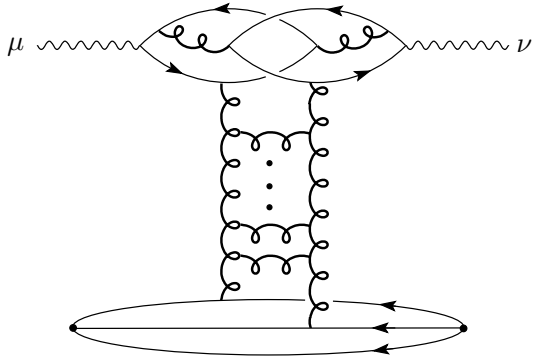}
\vspace*{.3cm}
\caption{Perturbative gluon ladder diagram which is contained in the 
generic diagram of figure \ref{fig2}b for the amplitude $\mathcal{M}^{(b)}$ 
\label{fig200b}}
\end{center}
\end{figure}

\begin{figure}[ht]
\begin{center}
\includegraphics[width=5.4cm]{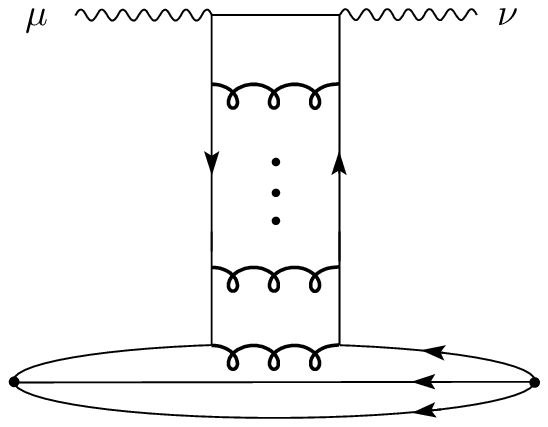}
\vspace*{.3cm}
\caption{Perturbative quark ladder diagram which is contained in the 
generic diagram of figure \ref{fig2}c for the amplitude $\mathcal{M}^{(c)}$ 
\label{fig201}}
\end{center}
\end{figure}

\begin{figure}[ht]
\begin{center}
\includegraphics[width=6.05cm]{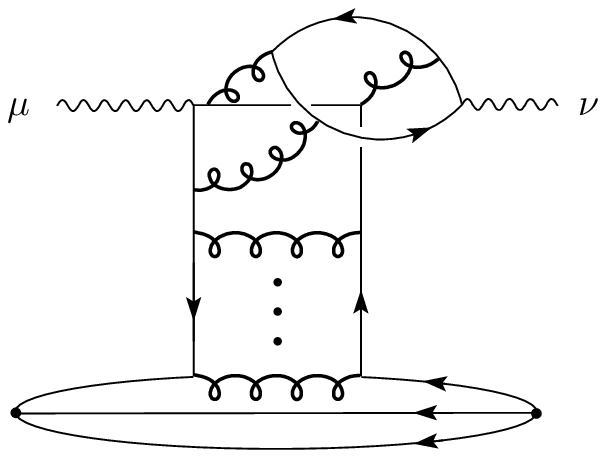}
\vspace*{.3cm}
\caption{Perturbative diagram which is contained in the 
generic diagram of figure \ref{fig2}g for the amplitude 
$\mathcal{M}^{(g)}$ 
\label{fig202}}
\end{center}
\end{figure}

Going one step further in sophistication we can have a look at 
resummed perturbation theory. In the perturbative framework the 
generic diagrams of figures \ref{fig2}a and \ref{fig2}b 
include gluon ladder diagrams, 
see figures \ref{fig200} and \ref{fig200b}, respectively. 
These are, of course, just the diagrams containing the leading order BFKL 
pomeron exchange \cite{Kuraev:fs,Balitsky:ic}. The leading perturbative 
diagrams contained in the generic diagrams 
of figures \ref{fig2}c-\ref{fig2}g, on the other hand, do not 
contain such gluon ladders. Instead, they typically contain quark-antiquark 
exchanges in the $t$-channel. As an example figure \ref{fig201} shows 
a perturbative quark ladder diagram contributing to the generic diagram 
of figure \ref{fig2}c. 
But it is well known that 
diagrams with quark-antiquark exchanges (or more generally 
fermion-antifermion exchanges) in the $t$-channel are suppressed 
by powers of the energy with respect to the corresponding diagrams 
containing gluon (or more generally vector boson) 
exchanges. Similarly, also the generic diagrams 
in figures \ref{fig2}d and \ref{fig2}e contain quark-antiquark 
exchanges as the leading terms in perturbation theory. 
Presumably also the leading terms of the generic diagrams of 
figures \ref{fig2}f and \ref{fig2}g correspond to quark-antiquark exchanges, 
see figure \ref{fig202},
where this is shown for the diagram of figure \ref{fig2}g. 
We can hence conclude that the leading perturbative diagrams 
for the generic diagrams of figures \ref{fig2}c-\ref{fig2}g are 
suppressed by powers of $\sqrt{s}$ with respect to the leading 
perturbative contributions to the amplitudes $\mathcal{M}^{(a)}$ 
and $\mathcal{M}^{(b)}$. 

We also arrive at the same conclusion when we switch from 
perturbation theory to Regge theory. 
Quite generally, we expect the dominant contribution from the 
diagrams of figures \ref{fig2}a and \ref{fig2}b to be due to pomeron 
exchange, soft and hard, with its well-known phenomenological 
properties, see for instance \cite{Donnachie:en}. 
On the other hand, we estimate that the main contribution from 
the generic diagram of figure \ref{fig2}c for large $\sqrt{s}$ is due 
to reggeon exchange which would correspond to the diagram
of figure \ref{fig201}, now interpreted as a diagram of Regge theory. 
Again, we know from phenomenology that reggeon exchanges are 
suppressed by a factor of roughly $1/\sqrt{s}$ relative to soft pomeron 
exchange, see \cite{Donnachie:en}. 
Thus we can argue that the generic diagrams of figure \ref{fig2}c 
and in a similar way those of figures \ref{fig2}d and \ref{fig2}e 
correspond to reggeon exchanges and are thus not leading 
for large $\sqrt{s}$. The same holds for the diagrams of 
figures \ref{fig2}f and \ref{fig2}g, as is illustrated in figure 
\ref{fig202}. Again we find that the amplitudes 
$\mathcal{M}^{(c)}$ - $\mathcal{M}^{(g)}$ are subleading 
at high energy $\sqrt{s}$. 

Let us now discuss the relative magnitudes of the amplitudes 
$\mathcal{M}^{(a)}$ and $\mathcal{M}^{(b)}$ corresponding 
to the figures \ref{fig200} and \ref{fig200b}. 
At high energies the leading contributions to both amplitudes 
have the same dependence on the energy. As discussed above 
that dependence results from Pomeron exchange or from 
gluon ladder diagrams, depending on whether one describes 
it in terms of Regge theory or perturbative QCD (pQCD). The 
typical perturbative diagrams are shown in figures 
\ref{fig200} and \ref{fig200b}, respectively, and in pQCD these diagrams 
represent the leading contribution at high energies. 
The main difference between the two amplitudes is the 
coupling of the gluon ladder (or the Pomeron) to the 
incoming and outgoing photon, or in perturbative 
language the impact factor. In the impact factor in 
$\mathcal{M}^{(b)}$ there are two quark loops which are 
connected by gluon lines. Due to the charge parity of the photon 
a minimum number of three gluons is required to couple to 
each of these loops, and the diagram shown in figure 
\ref{fig200b} is hence the lowest order contribution to the 
impact factor of $\mathcal{M}^{(b)}$. 
Compared with the amplitude $\mathcal{M}^{(a)}$ the 
amplitude $\mathcal{M}^{(b)}$ therefore contains two additional 
factors of $\alpha_s$ which are not compensated by any large 
logarithms. The relevant scale of $\alpha_s$ is determined by 
the photon virtuality $Q^2$. Consequently, the amplitude 
$\mathcal{M}^{(b)}$ is strongly suppressed at high virtualities. 
At small virtualities, however, $\alpha_s$ becomes large and 
that suppression is no longer relevant. Besides these factors 
of $\alpha_s$ there appears to be no other simple mechanism 
suppressing the diagrams of figure \ref{fig200b}. Accordingly, these 
diagrams can potentially give an important contribution in the region 
of low photon virtualities. In our opinion their contribution, which 
is often neglected in applications of the dipole picture of high 
energy scattering, deserves further study. 
It would be especially interesting to find ways of estimating the 
relative size of the terms $\mathcal{M}^{(a)}$ and 
$\mathcal{M}^{(b)}$ at low $Q^2$ using experimental data. 

It should be noted here that in our development of the dipole picture 
below we find that perturbative diagrams contained in figures 
\ref{fig2}a and \ref{fig2}b contribute to the colour dipole 
amplitude side and photon wave function side such that 
one can speak of a mixing of the diagram classes (a) and (b) 
in the colour dipole picture. This is relevant in 
higher orders in $\alpha_s$ as we will discuss when we make the 
transition to renormalised quantities in section 
\ref{sec:currenttoquarkqft} below. The preceding discussion 
refers to the leading terms (at large $\sqrt{s}$ and large $Q^2$) 
and is therefore not affected by this effect. 

From the observations made here we conclude that at 
high energies the amplitude $\mathcal{M}^{(a)}$ is 
expected to give the leading contribution if the photon 
virtuality is large, and that at lower virtualities also the 
amplitude $\mathcal{M}^{(b)}$ should become important. 
We will therefore in the following discuss 
in detail the amplitude $\mathcal{M}^{(a)}$, given in (\ref{2.8}), 
which corresponds to the diagram of figure \ref{fig2}a. For the amplitude 
$\mathcal{M}^{(b)}$, figure \ref{fig2}b, the same steps can be done 
in a completely analogous way, see appendix \ref{appC}. 

Finally we note that for the hypothetical case of scattering of an isovector 
photon on an $\Omega^-$ target only the diagram of figure \ref{fig2}a 
exists as long as one assumes equal up and down quark masses and 
neglects higher orders in the electromagnetic 
coupling $\alpha_{\rm em}$, as we do throughout the present papers. 
Indeed, in this case the current has the structure 
\be
J^\mu_{\rm isovector} = 
\frac{1}{2} \bar{u} \gamma^\mu u - \frac{1}{2} \bar{d}\gamma^\mu d \,,
\ee
whereas the $\Omega^-$ consists of three $s$-quarks. Thus none 
of the diagrams of figures \ref{fig2}b-\ref{fig2}g can be drawn for 
this case and $\mathcal{M}^{(a)}$  gives the complete amplitude. 

\subsection{From current to quark scattering for a fixed gluon potential}
\label{sec:currenttoquark}

In the following we consider the diagram of figure \ref{fig2}a. We will in essence cut the
quark lines at the vertex marked $x$, insert factors of $1$, and relate the amplitude
for the scattering of the photon on the proton to that of scattering of a
$q\bar{q}$ pair on the proton. We recall first that the free quark propagator for mass
$m_q$ satisfies\footnote{Note that the superscript $0$ in the quark propagator 
$S^{(q,0)}_F$ indicates that this is the free propagator, whereas in the 
quark mass $m_q^{(0)}$ it indicates that this is the bare mass, see (\ref{A.1}).}
\begin{eqnarray}
\label{2.11}
(i\rightslash_x-m_q)S_F^{(q,0)}(x,y)
&=&
{}
-\delta^{(4)}(x-y)\,,\\
S^{(q,0)}_F(x,y)(i\leftslash_y+m_q)
&=&
{}
\delta^{(4)}(x-y)\,. 
\label{2.12}
\end{eqnarray}
A simple exercise shows that $S_F^{(q,0)}$ can be written as a sum of dyadic 
products of quark and antiquark wave functions, see (\ref{B.6}) of 
appendix \ref{appB}.

\begin{figure}
\begin{center}
\includegraphics[width=9.6cm]{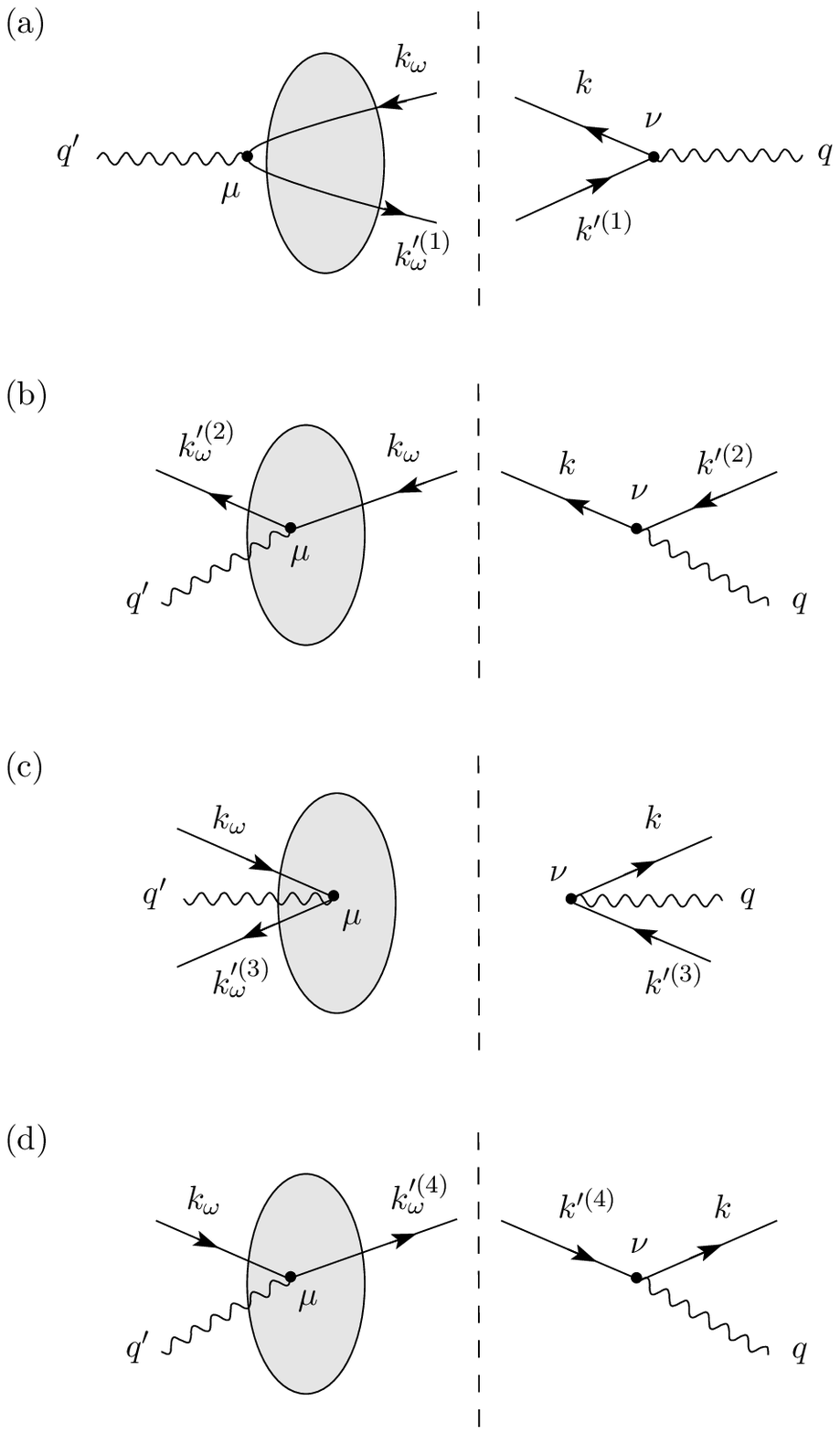}
\caption{Interpretation of the four terms $A_j^{(q)\mu\nu}(q)$ of (\ref{2.15})
where $j=1,\dots,4$ corresponds
to (a), \dots, (d). The shading indicates the fixed gluon potential. 
The arrows on the lines indicate the interpretation as quark or antiquark. 
The orientation of the momentum assignment to the lines is from 
right to left for all lines, irrespective of the arrows on the lines. 
\label{fig3}}
\end{center}
\end{figure}

We shall now use (\ref{2.11}), (\ref{2.12}) and (\ref{B.6}) to rewrite 
$A^{(q)\mu\nu}(q)$ of (\ref{2.8a}) in a form where we can make the 
transition from current scattering to $q\bar{q}$ scattering. 
Inserting factors of one in (\ref{2.8a}) we get
\begin{eqnarray}
\label{2.13}
A^{(q)\mu\nu}(q)&=&\int d^4x~d^4z~d^4z^{\prime} \,
\mathrm{Tr}\left[\gamma^{\mu}S^{(q)}_F(0,z;G)\delta^{(4)}(z-x)\right.
\nonumber\\
&&{}
\left. 
e^{-iqx}\gamma^{\nu}\delta^{(4)}(x-z^{\prime})S^{(q)}_F(z^{\prime},0;G)
\right]
\nonumber\\
&=&\int d^4x~d^4z~d^4z^{\prime} \,
\mathrm{Tr}
\left[\gamma^{\mu}S^{(q)}_F(0,z;G)(-i\rightslash_z+m_q)S^{(q,0)}_F(z,x)
\right.
\nonumber\\
&&{}
\left.
e^{-iqx}\gamma^{\nu}S^{(q,0)}_F(x,z^{\prime})(i\leftslash_{z^{\prime}}+m_q)
S^{(q)}_F(z^{\prime},0;G)\right] \,.
\end{eqnarray}
Integration by parts and use of (\ref{B.6}) leads us to a representation 
of $A^{(q)\mu\nu}(q)$ as a sum of four terms which we can interpret 
in terms of $q$ and $\bar{q}$ scattering.
The key step in using (\ref{B.6}) is a decomposition of the free propagator 
$S^{(q,0)}_F$ as a spin and colour sum 
over Dirac spinors, see (\ref{spinsumu}), (\ref{spinsumv}). 
Those spinors immediately lead to an interpretation 
in terms of quarks and antiquarks. The details are given in appendix \ref{appB}. 
The final result derived there is (see (\ref{B.19}))
\begin{equation}\label{2.14}
A^{(q)\mu\nu}(q)=\sum^4_{j=1}A^{(q)\mu\nu}_j(q)\,,
\end{equation}
where 
\begin{equation}
\label{2.15}
A^{(q)\mu\nu}_j(q)=\frac{1}{2\pi}\int \frac{d\omega}{\omega+i\epsilon}\int
\frac{d^3k}{(2\pi)^3 2k^0} \, a_j\,,
\end{equation}
with 
\begin{eqnarray}
\label{2.16}
a_1&=&
-\sum_{r,r^{\prime}}(\overline{k^{\prime(1)}_\omega,r^{\prime}}|B^{(q) \mu}|k_\omega,r)
\bar{u}_r(k)\gamma^{\nu}v_{r^{\prime}}(k^{{\prime}(1)})
\nonumber\\
&&{}
(q^0-k^0-k^{\prime(1)0}-\omega+i\epsilon)^{-1}(2k^{\prime(1)0})^{-1}\,,
\end{eqnarray}
\begin{eqnarray}
\label{2.17}
a_2 &=&
\sum_{r,r^{\prime}}(k^{\prime(2)}_\omega,r^{\prime}|B^{(q) \mu}|k_\omega,r)
\bar{u}_r(k)\gamma^{\nu}u_{r^{\prime}}(k^{{\prime}(2)})
\nonumber\\
&&{}
(-q^0+k^0-k^{\prime(2)0}+\omega+i\epsilon)^{-1}
(2k^{\prime(2)0})^{-1}\,,
\end{eqnarray}
\begin{eqnarray}
\label{2.17a}
a_3&=&
-\sum_{r,r^{\prime}}(k^{\prime(3)}_\omega,r^{\prime}|B^{(q) \mu}|\overline{k_\omega,r})
\bar{v}_r(k)\gamma^{\nu}u_{r^{\prime}}(k^{{\prime}(3)})\nonumber\\
&&{}
(-q^0-k^0-k^{\prime(3)0}-\omega+i\epsilon)^{-1}(2k^{\prime(3)0})^{-1}\,,
\end{eqnarray}
\begin{eqnarray}
\label{2.17b}
a_4&=&
\sum_{r,r^{\prime}}(\overline{k^{\prime(4)}_\omega,r^{\prime}}|
B^{(q) \mu}|\overline{k_\omega,r})
\bar{v}_r(k)\gamma^{\nu}v_{r^{\prime}}(k^{{\prime}(4)})
\nonumber\\
&&{}
(q^0+k^0-k^{\prime(4)0}+\omega+i\epsilon)^{-1}
(2k^{\prime(4)0})^{-1}\,.
\end{eqnarray}
The operator $B^{(q) \mu}$ is defined in (\ref{B.8}) and 
$k^{\prime(j)},k^{\prime(j)}_\omega$
in (\ref{B.20})-(\ref{B.23}). 
Here $r,r'$ are spin and colour indices with $r=(\lambda,A)$, 
$\lambda=\pm 1/2$ and $A=1,2,3$. 
The four terms are illustrated in figure \ref{fig3}. 
In that figure the 
dashed vertical line indicates an integration over $\omega$ and over the 
three-momentum $\mathbf{k}$ according to (\ref{2.15}). 
We emphasise that the vertical line does not imply that all quark 
and antiquark lines are set on-shell, see the discussion below. 
The arrows on the lines indicate the interpretation 
as quark or antiquark. 
The orientation of the momentum assignment to the lines is from 
right to left for all lines, irrespective of the arrows on the lines. 
The interpretation of $A^{(q)\mu\nu}_j(q)$ is now easily obtained
(see figure \ref{fig3}): 
\begin{itemize}
\item
The term $A^{(q)\mu\nu}_1(q)$ corresponds in essence to the splitting of the
initial real or virtual
photon into a $q\bar{q}$ pair with momenta $k,k^{\prime(1)}$. 
This is combined with the amplitude for a $q\bar{q}$ pair of momenta 
$k_\omega,k^{\prime(1)}_\omega$ scattering on the fixed gluon potential 
and recombining to the final real or virtual photon $q^{\prime}$ 
(figure \ref{fig3}a).
\item
The term $A^{(q)\mu\nu}_2(q)$ corresponds to a quark of momentum $k^{\prime(2)}$
absorbing the photon $q$ and going to a quark of momentum $k$. This is combined with the
amplitude for a quark $k_\omega$ shaking off a photon $q^{\prime}$ in the fixed gluon
potential and going to quark $k^{\prime(2)}_\omega$ (figure \ref{fig3}b).
\item
The term $A^{(q)\mu\nu}_3(q)$ corresponds to an antiquark $k$ 
and quark $k^{\prime(3)}$ annihilating by absorption of a photon $q$. 
This is combined with the amplitude for the creation of a quark 
$k^{\prime(3)}_\omega$, antiquark $k_\omega$ and photon $q^{\prime}$ 
in the fixed gluon potential (figure \ref{fig3}c).
\item
The term $A^{(q)\mu\nu}_4(q)$ corresponds to an antiquark $k$ absorbing 
a photon $q$ and going to an antiquark $k^{\prime(4)}$. This is combined 
with the amplitude for an antiquark $k^{\prime(4)}_\omega$ going to 
an antiquark $k_\omega$ with emission of a photon $q^{\prime}$ in the 
fixed gluon potential (figure \ref{fig3}d).
\end{itemize}

Several points are noteworthy here. First, we have nowhere 
made the assumption that free quarks exist. 
Second, the mass parameter $m_q$  in the above formulae is 
completely arbitrary. 
Further, the spinors $u$ and $v$ in the splitting of the photon $q$ into a  
quark-antiquark pair etc.\ and the spinors in the scattering amplitudes 
$\overline{(k^{\prime(1)}_\omega, r^{\prime}}|B^{(q) \mu}|k_\omega,r)$ 
etc.\ are always those of quarks on the mass shell $m_q$, see (\ref{B.4}). 
But in the plane wave factors on the r.h.s.\ of (\ref{B.4}) the energies 
$k^0_\omega,k^{\prime(1)}_\omega$ etc.\ are not equal to the energies 
of on-shell quarks, see (\ref{B.3}) and (\ref{B.20})-(\ref{B.23}). 
For the diagrams in figure \ref{fig3} this means that the quark and 
antiquark lines to the right of the vertical line have energies corresponding 
to on-shell quarks. The quark and antiquark lines to the left of the 
vertical line, on the other hand, have energies corresponding to off-shell 
quarks, and these energies contain the integration parameter $\omega$. 
In this respect our result is reminiscent of noncovariant perturbation theory. 
But we stress that no perturbation expansion has been made in our 
approach. 

We should add a remark here concerning the interpretation 
of the different terms in figure \ref{fig3}. On first sight the occurrence 
of the last three diagrams might appear counterintuitive. 
We emphasise that all diagrams are obtained from the term 
$\mathcal{M}^{(a)}$ corresponding to figure \ref{fig2}a. 
The decomposition into the four terms of figure \ref{fig3} originates 
from the decomposition of propagators into spin sums. In that sense 
the free propagator contains all possible quark and antiquark spinor 
states, giving rise to the different combinations of incoming and 
outgoing quarks and antiquarks in the four terms in figure \ref{fig3}. 
Also that decomposition reminds one of noncovariant perturbation theory. 
Loosely speaking, the quark loop of the upper part of figure \ref{fig2}a 
is in figure \ref{fig3} cut and deformed 
in such a way that the intermediate quark and antiquark are `bent' 
into the initial or final state. 

\subsection {From current to quark scattering in QFT}
\label{sec:currenttoquarkqft}

Now we are ready to insert our result for $A^{(q)\mu\nu}(q)$ 
(\ref{2.14})-(\ref{2.17b}) in (\ref{2.8}) and to analyse the resulting 
expression. We get
\begin{equation}\label{2.20}
\mathcal{M}^{(a)\mu\nu}_{s^{\prime}s}(p^{\prime},p,q)=\sum^4_{j=1}
\mathcal{M}^{(a,j)\mu\nu}_{s^{\prime}s}(p^{\prime},p,q) 
\,,
\end{equation}
\begin{eqnarray}\label{2.21}
\mathcal{M}^{(a,j)\mu\nu}_{s^{\prime}s}(p^{\prime},p,q) &=&
-\frac{i}{2\pi m_p Z_p}\int d^4y^{\prime} \, d^4y \,
e^{ip^{\prime}y^{\prime}}\bar{u}_{s'}(p^{\prime})(-i\rightslash_{y'}+m_p)
\\
&&{}
\sum_q Q^2_q\Big\langle
\wick{2}{<1\psi_p(y^{\prime})>1{\overline{\psi}}_p(y)}
A^{(q)\mu\nu}_j(q)\Big\rangle_G \, (i\leftslash_y+m_p)u_s(p)e^{-ipy} \,,
\nonumber
\end{eqnarray}
\begin{eqnarray}\label{2.22}
\mathcal{M}^{(a,1)\mu\nu}_{s^{\prime}s}(p^{\prime},p,q)
\!&=&\!
\frac{1}{2\pi}\sum_qQ^2_q
\int\frac{d\omega}{\omega+i\epsilon}
\int\frac{d^3k}{(2\pi)^3 2k^0}
(q^0-k^0-k^{{\prime}(1)0}-\omega+i\epsilon)^{-1}
\nonumber\\
&&{}
(2k^{{\prime}(1)0})^{-1}
\sum_{r^{\prime},r}
\langle \gamma(q^{\prime},\mu), p(p^{\prime},s^{\prime})
|\mathcal{T}^{(a)}|\bar{q}(k^{{\prime}(1)}_\omega,r^{\prime}), 
q(k_\omega,r),p(p,s)\rangle\nonumber\\
&&{}
\bar{u}_r(k)Z_q\gamma^{\nu}v_{r^{\prime}}(k^{{\prime}(1)})\,,
\end{eqnarray}
\begin{eqnarray}\label{2.23}
\mathcal{M}^{(a,2)\mu\nu}_{s^{\prime}s}(p^{\prime},p,q)
\!&=&\!
\frac{1}{2\pi}\sum_q Q^2_q
\int\frac{d\omega}{\omega+i\epsilon}
\int\frac{d^3k}{(2\pi)^32k^0}
(-q^0+k^0-k^{{\prime}(2)0}+\omega+i\epsilon)^{-1}
\nonumber\\
&&{}
(2k^{{\prime}(2)0})^{-1}
\sum_{r^{\prime},r}\langle 
\gamma(q^{\prime},\mu),q(k^{{\prime}(2)}_\omega,r^{\prime}),
p(p^{\prime},s^{\prime})|\mathcal{T}^{(a)}|q(k_\omega,r),p(p,s)
\rangle\nonumber\\
&&{}
\bar{u}_r(k)Z_q\gamma^{\nu}u_{r^{\prime}}(k^{{\prime}(2)})\,,
\end{eqnarray}
\begin{eqnarray}\label{2.24}
\mathcal{M}^{(a,3)\mu\nu}_{s^{\prime}s}(p^{\prime},p,q)
\!&=&\!
\frac{1}{2\pi}\sum_qQ^2_q
\int\frac{d\omega}{\omega+i\epsilon}
\int\frac{d^3k}{(2\pi)^32k^0}
(-q^0-k^0-k^{{\prime}(3)0}-\omega+i\epsilon)^{-1}
\nonumber\\
&&{}
(2k^{{\prime}(3)0})^{-1}
\sum_{r^{\prime},r}\langle 
\gamma(q^{\prime},\mu), 
\bar{q}(k_\omega,r), q(k^{{\prime}(3)}_\omega,r^{\prime}),
p(p^{\prime},s^{\prime})|\mathcal{T}^{(a)}|p(p,s)\rangle\nonumber\\
&&{}
\bar{v}_r(k)Z_q\gamma^{\nu}u_{r^{\prime}}(k^{{\prime}(3)})\,,
\end{eqnarray}
\begin{eqnarray}\label{2.25}
\mathcal{M}^{(a,4)\mu\nu}_{s^{\prime}s}(p^{\prime},p,q)
\!&=&\!
\frac{1}{2\pi}\sum_qQ^2_q
\int\frac{d\omega}{\omega+i\epsilon}
\int\frac{d^3k}{(2\pi)^32k^0}
(q^0+k^0-k^{{\prime}(4)0}+\omega+i\epsilon)^{-1}
\nonumber\\
&&{}
(2k^{{\prime}(4)0})^{-1}
\sum_{r^{\prime},r}\langle 
\gamma(q^{\prime},\mu),
\bar{q}(k_\omega,r), 
p(p^{\prime},s^{\prime})|\mathcal{T}^{(a)}|\bar{q}(k^{{\prime}(4)}_\omega,
r^{\prime}),p(p,s)\rangle\nonumber\\
&&{}
(-1) \bar{v}_r(k)Z_q\gamma^{\nu}v_{r^{\prime}}(k^{{\prime}(4)})\,, 
\end{eqnarray}
where we have defined 
\begin{eqnarray}\label{2.26}
\lefteqn{
\langle \gamma(q^{\prime},\mu),p(p^{\prime},s^{\prime})|\mathcal{T}^{(a)}|
\bar{q}(k^{{\prime}(1)}_\omega,r^{\prime}),q(k_\omega,r),p(p,s)\rangle =
}
\hspace{1cm}
\\
&=&{}
-\frac{i}{2\pi m_p Z_pZ_q}\int d^4y^{\prime}\,d^4y\, e^{ip^{\prime}y^{\prime}}
\bar{u}_{s^{\prime}}(p^{\prime})(-i\rightslash_{y'}+m_p)\nonumber\\
&&{}
\Big\langle \wick{2}{<1\psi_p(y^{\prime})>1{\overline{\psi}}_p(y)(-1)
(\overline{k_\omega^{{\prime}(1)},r^{\prime}}|B^{(q) \mu}|k_\omega,r)\Big\rangle_G \,
(i\leftslash_y+m_p)u_s(p)e^{-ipy}}\,,
\nonumber \\
\label{2.27}
\lefteqn{
\langle \gamma(q^{\prime},\mu), 
q(k^{{\prime}(2)}_\omega,r^{\prime}),
p(p^{\prime},s^{\prime})|\mathcal{T}^{(a)}|q(k_\omega,r),p(p,s)\rangle =
}
\hspace{1cm}
\\
&=&{}
-\frac{i}{2\pi m_pZ_pZ_q}\int d^4y^{\prime}\,d^4y\,e^{ip^{\prime}y^{\prime}}\bar{u}_{s^{\prime}}
(p^{\prime})(-i\rightslash_{y'}+m_p)\nonumber\\
&&{}
\Big\langle\wick{2}{<1 \psi_p(y^{\prime})>1{\overline{\psi}}_p(y)
(k^{{\prime}(2)}_\omega,r^{\prime}|B^{(q) \mu}|k_\omega,r)\Big\rangle_G \,
(i\leftslash_y+m_p)u_s(p)e^{-ipy}}\,,
\nonumber \\
\label{2.28}
\lefteqn{
\langle \gamma(q^{\prime},\mu),\bar{q}(k_\omega,r),
q(k^{{\prime}(3)}_\omega,r^{\prime}),
p(p^{\prime},s^{\prime})|\mathcal{T}^{(a)}|p(p,s)\rangle =
}
\hspace{1cm}
\\
&=&{}
-\frac{i}{2\pi m_pZ_pZ_q}\int d^4y^{\prime}\,d^4y \,
e^{ip^{\prime}y^{\prime}}\bar{u}_{s^{\prime}}
(p^{\prime})(-i\rightslash_{y'}+m_p)
\nonumber\\
&&{}
\Big\langle\wick{2}{<1\psi_p(y^{\prime})>1{\overline{\psi}}_p(y)(-1)
(k^{{\prime}(3)}_\omega,r^{\prime}|B^{(q) \mu}|\overline{k_\omega,r})\Big\rangle_G\,
(i\leftslash_y+m_p)u_s(p)e^{-ipy}}\,,
\nonumber\\
\label{2.29}
\lefteqn{
\langle \gamma(q^{\prime},\mu),
\bar{q}(k_\omega,r),p(p^{\prime},s^{\prime})|\mathcal{T}^{(a)}|\bar{q}
(k^{{\prime}(4)}_\omega,r^{\prime}),p(p,s)\rangle =
}
\hspace{1cm}
\\
&=&{}
-\frac{i}{2\pi m_pZ_pZ_q}\int d^4y^{\prime}\,d^4y \,
e^{ip^{\prime}y^{\prime}}\bar{u}_{s^{\prime}}
(p^{\prime})(-i\rightslash_{y^{\prime}}+m_p)
\nonumber\\
&&{}
\Big\langle\wick{2}{<1\psi_p(y^{\prime})>1{\overline{\psi}}_p(y)
(-1) (\overline{k^{{\prime}(4)}_\omega,r^{\prime}}|B^{(q) \mu}|
\overline{k_\omega,r})\Big\rangle_G\,
(i\leftslash_y+m_p)u_s(p)e^{-ipy}}\,.
\nonumber
\end{eqnarray}

Here it seems that we can already make a connection to the usual 
dipole picture for photon-induced reactions. The amplitude 
$\mathcal{M}^{(a,1)}$ (\ref{2.22}) has been written as a product 
of a term describing the splitting of the initial photon $\gamma(q,\nu)$ 
into quark and antiquark and the subsequent scattering of the quark 
and antiquark off the proton to give the final state 
$|p(p^{\prime},s^{\prime}),\gamma(q^{\prime},\mu)\rangle$. 
But there are two obvious complications.

\begin{figure}
\begin{center}
\includegraphics[width=9.4cm]{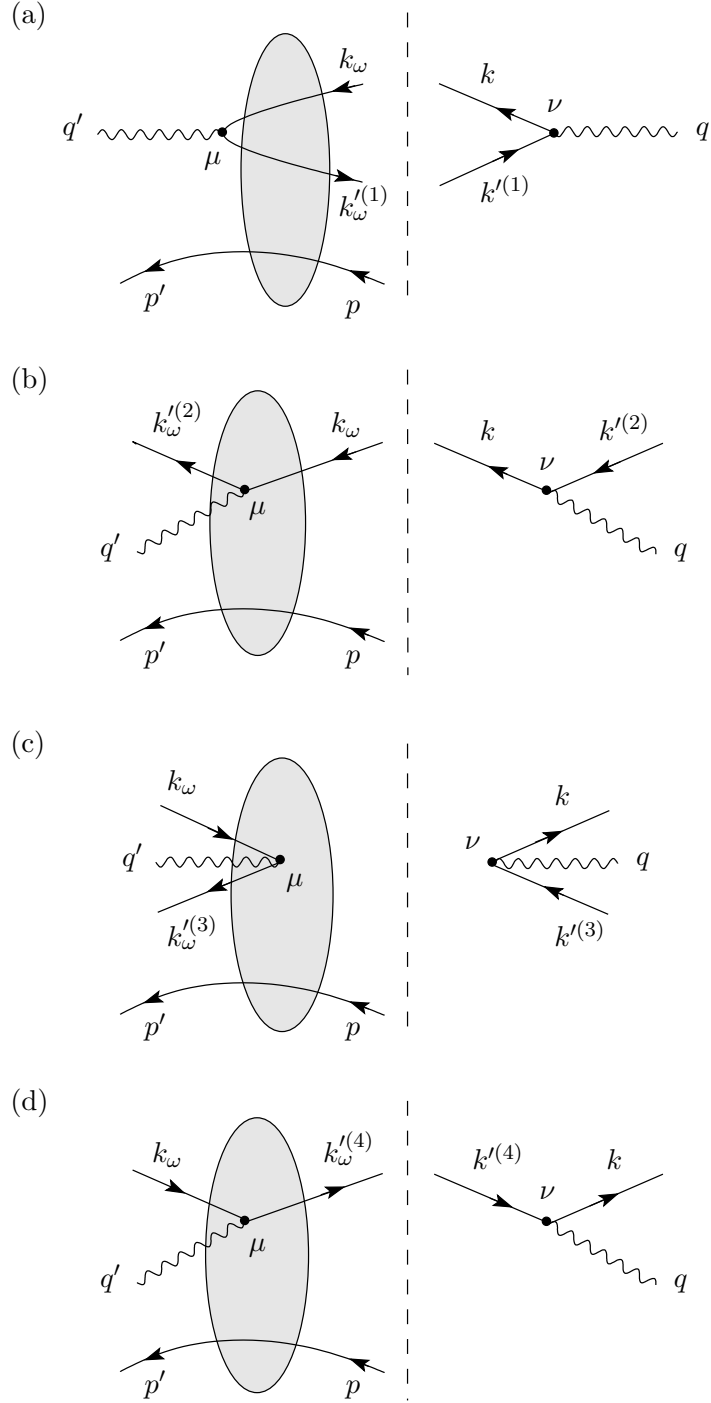}
\caption{Interpretation of the four terms $\mathcal{M}^{(a,j)}_{s^\prime s}(p^\prime, p,q)$
(\ref{2.22})-(\ref{2.25}) where $j=1,\dots,4$ corresponds to (a),\dots,(d). The shading
indicates all possible interactions of the proton with the quarks at the top of the
diagrams. 
The arrows on the lines indicate the interpretation as particle or antiparticle. 
The orientation of the momentum assignment to the lines is from 
right to left for all lines, irrespective of the arrows on the lines. 
\label{fig4}}
\end{center}
\end{figure}

First, we have written $\mathcal{M}^{(a)}$ as a sum of four terms in (\ref{2.20}). 
The interpretation of these terms is easily obtained from 
(\ref{2.22})-(\ref{2.25}) and is shown in figures \ref{fig4}a-\ref{fig4}d. 
It is analogous to the interpretation of the four diagrams of figure $\ref{fig3}$ 
in the preceding section. 
The vertex diagrams on the r.h.s.\ of figure \ref{fig4} correspond to the 
vertex factors $\bar{u}_r (k) Z_q \gamma^\nu v_{r'} (k^{\prime (1)})$ 
etc.\ in (\ref{2.22})-(\ref{2.25}). The diagrams on the l.h.s.\ of figure 
\ref{fig4} correspond to the matrix elements (\ref{2.26})-(\ref{2.29}). 
We emphasise that these are matrix elements for quarks off the 
energy shell. The normalisations in (\ref{2.26})-(\ref{2.29}) are, however, 
chosen with correct LSZ factors and wave function renormalisation 
constants in such a way that the matrix elements of $\mathcal{T}^{(a)}$ 
are finite and would correspond to physical $T$-matrix elements 
under the following conditions: 
\begin{itemize}
\item
The quarks $q$ are supposed to have a mass shell with a pole mass 
$m_{q,{\mathrm{pole}}}$ and we choose our mass 
parameter -- which is still at our free disposal -- equal to the 
pole mass. That is, we set $m_q= m_{q,{\mathrm{pole}}}$. 
\item
We replace the momenta of the quarks and antiquarks in 
(\ref{2.26})-(\ref{2.29}) -- which are in general off the energy shell -- 
by the on-shell momenta. That is we make the replacement 
$k_\omega \to k$, $k_\omega^{(1)} \to k^{(1)}$ etc. 
\end{itemize}
Keeping these points in mind we see that 
figure \ref{fig4}a is clearly interpretable in the dipole picture, in 
contrast to the other three diagrams. 
Figure \ref{fig4}b describes the absorption of the 
photon $q$ on a quark $k^{{\prime}(2)}$ going to quark $k$. This is 
combined with the amplitude  for scattering of quark $k_\omega$ on 
the proton giving quark $k^{{\prime}(2)}_\omega$, a photon $q^{\prime}$ 
and the proton $p^{\prime}$. The interpretation of the figures \ref{fig4}c 
and \ref{fig4}d is analogous.

A second complication arises when we consider the renormalisation of the 
photon vertex. 
We have inserted a wave function renormalisation factor $Z^{-1}_q$ 
in the scattering amplitudes $\mathcal{T}^{(a)}$ (\ref{2.26})-(\ref{2.29}), 
since otherwise they would not be finite. Correspondingly an explicit factor 
$Z_q$ has been put into the splitting function of the initial photon 
$\gamma(q,\nu)$.
Thus we are still dealing with (in general) divergent integrands in (\ref{2.22})-(\ref{2.25})
due to the explicit factor $Z_q$. This divergence in $Z_q$ must, however, be cancelled by
a divergence in the integral over $\omega$ and $\mathbf{k}$.

\begin{figure}[b]
\begin{center}
\includegraphics[width=11.4cm]{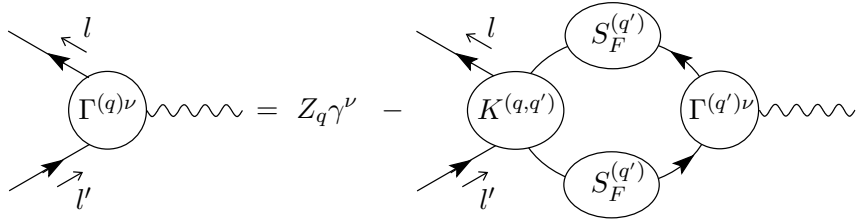}
\caption{The equation satisfied by the renormalised $\gamma q\bar{q}$ vertex
function in diagrammatic form. In the last term a sum over the quark flavour 
$q'$ (see eq.\ (\ref{2.30})) is implied. 
The arrows on the lines indicate the interpretation as quark or antiquark. 
\label{fig5}}
\end{center}
\end{figure}

The situation here is similar to the case of overlapping divergences in 
QED.  The way to deal with such overlapping divergences is well known 
(see for instance \cite{5}). Let $\Gamma^{(q)\nu}(l,l')$ be 
the renormalised $\gamma q\bar{q}$ vertex function. 
This function satisfies a Dyson-Schwinger equation shown diagrammatically 
in figure \ref{fig5}, which is obtained as a straightforward generalisation 
of the corresponding equation in QED, see \cite{5}. 
Here $K^{(q,q')}$ is the renormalised kernel of $q'\bar{q}'$ to 
$q\bar{q}$ scattering which contains all connected Feynman diagrams 
which are quark-antiquark two particle irreducible in the $s$-channel 
and $S^{(q')}_F$ is the renormalised quark propagator. 
Since we neglect all higher order terms in the electromagnetic coupling 
$\alpha_{\rm em}$ the diagram expansion of the kernel comprises only QCD 
diagrams. Among the diagrams contained in $K^{(q,q')}$ is the 
one-gluon exchange between the quark and antiquark, as well 
as the annihilation of the quark and antiquark into three gluons 
in the $s$-channel which then create another quark-antiquark 
pair (compare the diagram in figure \ref{figinterab} below). 
In the latter type of diagram the photon can couple to a $q\bar{q}$ 
pair of a different flavour than the one on the left. This possibility requires 
the two flavour indices in the notation $K^{(q,q')}$. 

\begin{figure}[ht]
\begin{center}
\includegraphics[width=10.95cm]{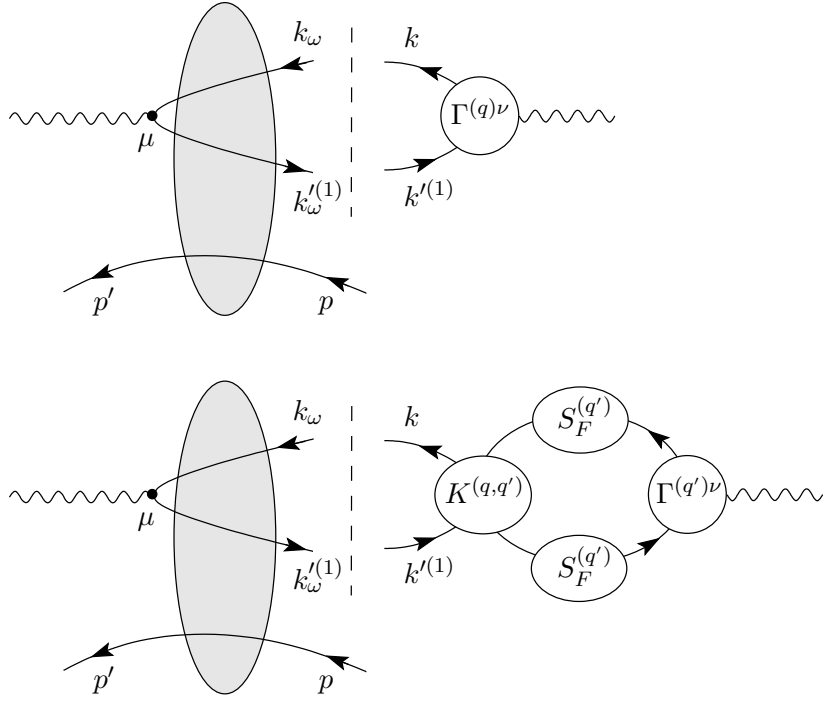}
\caption{The two terms for the amplitude $\mathcal{M}^{(a,1)}$ (\ref{2.33})
in diagrammatic form. Again a sum over the quark flavours $q'$ is implied 
in the second term. 
The arrows on the lines indicate the interpretation as particle or antiparticle. 
The orientation of the momentum assignment to the lines is from 
right to left, irrespective of the arrows on the lines. 
\label{fig6}}
\end{center}
\end{figure}

In a shorthand notation the Dyson-Schwinger equation reads 
\begin{equation}\label{2.30}
\Gamma^{(q)\nu} (l,l') =
Z_q\gamma^{\nu}-
\sum_{q'}\int K^{(q,q')}S^{(q')}_F \Gamma^{(q')\nu}S^{(q')}_F\,. 
\end{equation}
Solving for $Z_q\gamma^{\nu}$ we get 
\begin{equation}\label{2.31}
Z_q\gamma^{\nu}
=\Gamma^{(q)\nu}(l,l^{\prime})
 + \sum_{q'}\int K^{(q,q')}S^{(q')}_F \Gamma^{(q')\nu} S^{(q')}_F
\,.
\end{equation}
All quantities on the r.h.s.\ of this equation are renormalised quantities, 
in particular, they are finite. The divergence of $Z_q$ originates 
from the integration in the last term. 
Replacing now $Z_q\gamma^{\nu}$ in (\ref{2.22})-(\ref{2.25}) according 
to (\ref{2.31}) we get expressions for $\mathcal{M}^{(a,j)}$ for $j=1,\ldots,4$ 
where all integrands are finite. We still have some freedom how to choose 
the momenta $l,l^{\prime}$ on the r.h.s.\ of (\ref{2.31}). A natural choice 
for $\mathcal{M}^{(a,1)}$ is
\begin{equation}\label{2.32}
l=k,~l^{\prime}=-k^{{\prime}(1)}\,.
\end{equation}
With this the final result for $\mathcal{M}^{(a,1)}$ is shown in diagrammatic 
form in figure \ref{fig6} 
and reads
\begin{eqnarray}\label{2.33}
\mathcal{M}^{(a,1)\mu\nu}_{s^{\prime}s}(p^{\prime},p,q)
\!&=&\!
\frac{1}{2\pi} \sum_qQ^2_q
\int\frac{d\omega}{\omega+i\epsilon}\int\frac{d^3k}{(2\pi)^32k^0}
(q^0-k^0-k^{\prime(1)0}-\omega+i\epsilon)^{-1}
\nn \\
&&{} 
(2k^{\prime(1)0})^{-1}
\sum_{r^{\prime},r}\langle \gamma(q^{\prime},\mu),
p(p^{\prime},s^{\prime})|\mathcal{T}^{(a)}|\bar{q}(k^{\prime(1)}_\omega,r^{\prime}),
q(k_\omega,r),p(p,s)\rangle
\nonumber\\
&&{}
\bar{u}_r(k)\left\{\Gamma^{(q)\nu}(k,-k^{\prime(1)})
+ \sum_{q'}\int K^{(q,q')}S^{(q')}_F\Gamma^{(q')\nu}S^{(q')}_F\right\}
v_{r^{\prime}}(k^{\prime(1)})\,.
\nn \\
&&{}
\end{eqnarray}
The explicit expressions and diagrammatic forms for 
$\mathcal{M}^{(a,j)}, j=2,3,4$ are analogous and easily constructed. 
In II we shall study the high energy limit, where we will find that
$\mathcal{M}^{(a,1)}$ gives the leading term.

The integrations over $\omega$ and $\mathbf{k}$ (\ref{2.33}) 
should be convergent and 
finite, but only for both terms in the curly brackets inserted together. 
If we insert for instance only $\Gamma^{(q)\nu}$ a divergence could occur. 
It would be very interesting to study in detail how the divergences 
cancel in this equation. It is likely that already a one-loop calculation 
would reveal the structure of this cancellation, possibly even in a simple 
model like the abelian gluon model proposed in \cite{Nachtmann:2000gm}. 

\begin{figure}[ht]
\begin{center}
\includegraphics[width=11.9cm]{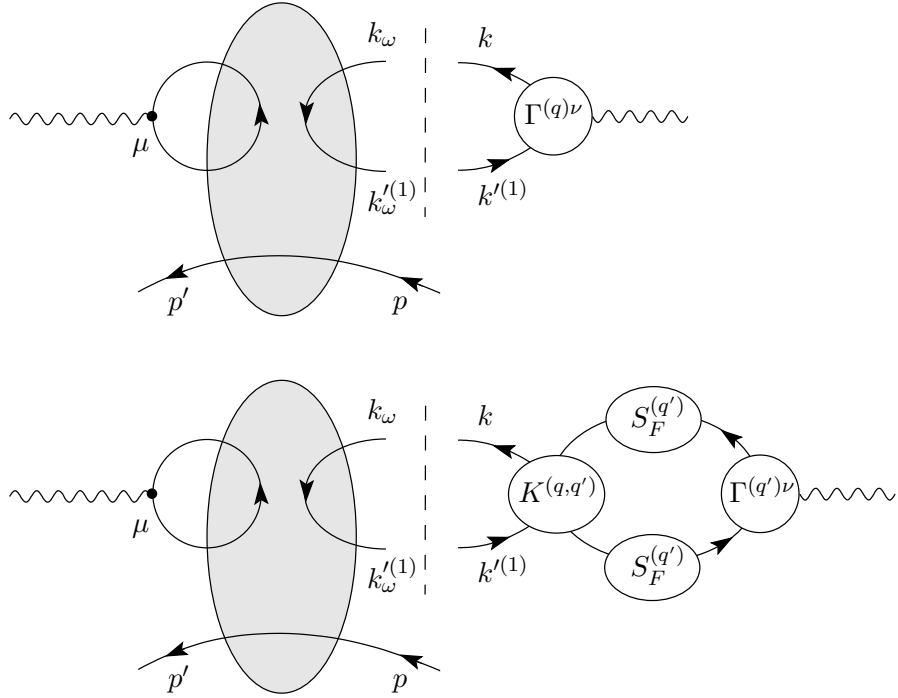}
\caption{The two terms for the amplitude $\mathcal{M}^{(b,1)}$ (\ref{2.45})
in diagrammatic form. The interpretation of the lines 
is as in figure \protect\ref{fig6}. 
\label{fig9}}
\end{center}
\end{figure}

For the amplitude ${\cal M}^{(b)}$ corresponding to the generic diagram 
of figure \ref{fig2}b all the same steps can be performed as for 
${\cal M}^{(a)}$. We get again a decomposition as in (\ref{2.20}), 
\be
\label{2.44}
\mathcal {M}^{(b) \mu \nu}_{s's} (p',p,q) = 
\sum_{j=1}^4 \mathcal {M}^{(b,j) \mu \nu}_{s's} (p',p,q) 
\,,
\ee
where of these four terms the important term at high energies is 
\bea
\label{2.45}
\mathcal {M}^{(b,1) \mu \nu}_{s's} (p',p,q)\! &=&\! 
\frac{1}{2\pi} \sum_{q'',q} Q_{q''} Q_q 
\int\frac{d\omega}{\omega+i\epsilon}\int\frac{d^3k}{(2\pi)^3 2k^0}
(q^0-k^0-k^{\prime(1)0}-\omega+i\epsilon)^{-1}
\nn \\
&&{}\!
(2k^{\prime(1)0})^{-1}
\sum_{r^{\prime},r}\langle \gamma(q^{\prime},\mu),
p(p^{\prime},s^{\prime})|\mathcal{T}^{(q'',b)}|\bar{q}(k^{\prime(1)}_\omega,r^{\prime}),
q(k_\omega,r),p(p,s)\rangle
\nn \\
&&{}\!
\bar{u}_r(k)\left\{\Gamma^{(q)\nu}(k,-k^{\prime(1)})
+ \sum_{q'}\int K^{(q,q')}S^{(q')}_F\Gamma^{(q')\nu}S^{(q')}_F\right\}
v_{r^{\prime}}(k^{\prime(1)})\,.
\nn \\
\eea
Graphically this is represented in figure \ref{fig9}. 
The explicit expressions are given in appendix \ref{appC}. The matrix 
element $\langle \gamma, p |\mathcal{T}^{(q'',b)}| \bar{q}, q, p \rangle$ 
in (\ref{2.45}) represents the scattering of a $q\bar{q}$ dipole on a proton 
giving a photon component $\bar{q}'' \gamma^\mu q''$ and a proton. 

\begin{figure}
\begin{center}
\includegraphics[width=6.55cm]{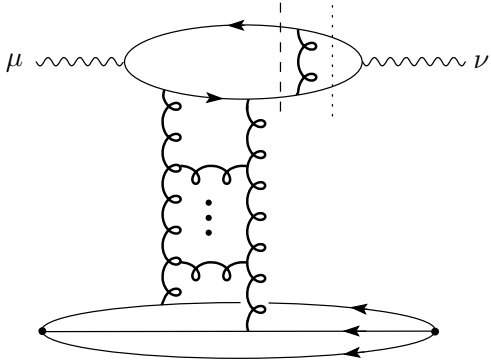}
\caption{Perturbative diagram contributing to the amplitude $\mathcal{M}^{(a,1)}$ 
  \label{figvirtab}}
\end{center}
\end{figure}

It is instructive to illustrate the overlapping divergences problem which 
is solved by (\ref{2.33}) and (\ref{2.45}) with two simple perturbative 
diagrams. Consider first the diagram of figure \ref{figvirtab}. 
Cutting this diagram along the dashed line gives a higher order correction 
to the photon vertex function. Cutting it along the dotted line we find a 
higher order correction to the scattering amplitude of the colour dipole 
on the proton. In the expression (\ref{2.33}) both corrections are taken 
into account and the `double counting' problem is exactly taken care of 
by the kernel term $K^{(q,q')}$. 

This double counting or overlapping divergences problem 
in the `cut' diagrams and its solution in (\ref{2.33}) and (\ref{2.45}) 
also `mixes' the diagrams of figures \ref{fig2}a and \ref{fig2}b. 
This is despite the fact that these diagram classes are clearly 
identifiable and do not mix at the starting level. Indeed, consider 
figure \ref{figinterab}, which is clearly a diagram of type \ref{fig2}b, 
and cut the diagram first along the dotted line. 
\begin{figure}[ht]
\begin{center}
\includegraphics[width=7.85cm]{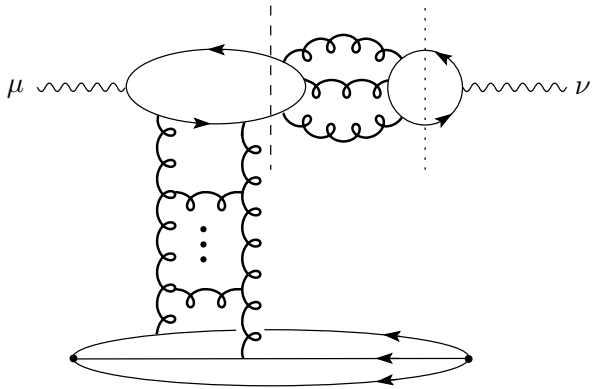}
\caption{Perturbative diagram contributing to the amplitude 
$\mathcal{M}^{(a,1)}$ but also to $\mathcal{M}^{(b,1)}$ 
\label{figinterab}}
\end{center}
\end{figure}
Then we have a colour dipole-proton amplitude of type 
\ref{fig2}b with a simple vertex on the photon side. 
Cutting along the dashed line, 
we can interpret it as a colour dipole-proton amplitude 
of type \ref{fig2}a with a vertex correction on the photon side. 
This shows that great care has to be exercised when 
cutting perturbative diagrams and trying to interpret them in 
terms of colour dipole amplitude times photon wave function. 
We would like to stress that the original classification of 
diagram types in figure \ref{fig2} is not affected by this. 
The `mixing' of diagram classes comes into play only when 
one considers cut diagrams, and then an individual 
diagram like the one in figure \ref{figinterab} 
can be associated with higher order corrections occurring 
to the left or to the right from the cut, depending on the 
position of the cut. 

Finally we want to make some remarks concerning 
the outgoing photon. 
As we indicate briefly in appendix \ref{appA} one can treat the 
outgoing photon $\gamma(q',\mu)$ of the Compton amplitude 
in complete analogy to the incoming photon $\gamma(q,\nu)$, 
performing all the steps we have done for the 
incoming photon in the same manner, including the 
renormalisation of the the photon vertex via its Dyson-Schwinger 
equation. The resulting expressions would in their form 
be completely symmetric in the incoming and outgoing photon. 
For the purpose of the present papers, especially for the study 
of the high energy limit to be discussed in II, it will however 
be sufficient to treat only the incoming photon in this way. 

\section{Summary}
\label{sumIsect}

In the present paper we have studied the real and virtual Compton 
amplitude for photon-nucleon scattering in a nonperturbative 
framework. We have classified the different contributions 
to the amplitude according to their quark line skeleton. 
We have discussed the relative magnitude of the different 
contributions and have identified the two contributions 
$\mathcal{M}^{(a)}$ and $\mathcal{M}^{(b)}$ 
represented by figures \ref{fig2}a and \ref{fig2}b as the 
leading ones at high energies. Concentrating on these two 
contributions we have then inserted on the quark lines 
after the incoming photon vertex suitable factors of one, 
given by the free Dirac operator acting on the free quark 
propagator for a chosen mass $m_q$. After integrating by parts 
we have used spin times colour sum decompositions 
for the free quark 
propagators to obtain expressions which bring 
us closer to the dipole picture of high energy scattering. 
For both amplitudes 
$\mathcal{M}^{(a)}$ and $\mathcal{M}^{(b)}$ we obtained 
four terms where in each case only one had the structure 
of a vertex function describing the splitting of the photon 
into a $q\bar{q}$ pair convoluted with a scattering amplitude 
for an off energy shell $q \bar{q}$ pair on the proton. The results 
for these parts of the amplitudes $\mathcal{M}^{(a)}$ and 
$\mathcal{M}^{(b)}$ are given in (\ref{2.33}) and (\ref{2.45}), 
respectively, see figures \ref{fig6} and \ref{fig9}. 
In II these results will allow us to isolate 
the leading term in the high energy limit. A key point in the 
present paper is that we have taken into account the 
renormalisation of the photon-quark-antiquark vertex. 
We find that the renormalisation induces an important 
rescattering correction which is usually not taken into account 
in the dipole picture. 

From the contribution 
of figure \ref{fig2}a we will be able to obtain the usual dipole 
picture of photon-proton scattering after suitable further 
approximations. We will in fact show that this is the leading term 
at high energies and large photon virtualities. 
We find, however, that there are two 
important correction terms which can potentially become 
large at small photon virtualities, that is in the interesting 
transition region from the perturbative to the nonperturbative 
regime. The first correction comes from the rescattering of 
the quark and antiquark into which the virtual photon splits, 
the second correction comes from the contribution of 
figure \ref{fig2}b. Both of them are suppressed only at large 
photon virtualities but not due to large energies. 

The motivation for the dipole picture of high energy scattering 
is closely related to the notion of light-cone wave functions, 
see for example \cite{Lepage:1979zb,Lepage:1980fj}. 
When applied to a real or virtual photon this concept naturally 
suggests to include in addition to the quark-antiquark 
component 
also higher Fock states of the photon like 
for example the quark-antiquark-gluon component 
of the wave function etc., and to sum the amplitudes for 
the scattering of these components off the proton. 
Also the perturbative framework can be arranged according 
to this view. We emphasise that in our treatment nothing 
is missing and, in particular, higher Fock states do not 
and should not occur explicitly here. 
Recall in this context that the solution of the overlapping 
divergences problem for the vacuum polarisation in QED 
(see \cite{5}) also does not involve analogues of higher 
Fock states for the photon. Nevertheless we think that 
it is an interesting problem for further investigations 
to try to see where and how the various terms of a Fock state 
expansion of the photon as discussed for instance in 
\cite{Hebecker:1999ej} can be identified in our approach. 

In the second paper we will discuss in detail the high energy 
limit of the expressions (\ref{2.33}) and (\ref{2.45}) obtained here 
from the contributions of figures \ref{fig2}a and \ref{fig2}b. 
We will also study the problem of gauge invariance. In particular, we will 
discuss whether different parts of the amplitude are 
separately gauge invariant. We point out the consequences of this 
issue for the correct definition of the perturbative wave function of 
the photon. Finally we derive some phenomenological consequences 
of the simple dipole picture which can help to find its range of validity. 
With this brief outlook we want 
to close the present paper, further conclusions will be presented 
in the second paper. 

\section*{Acknowledgements}

We would like to thank W.\ Buchm\"uller, 
M.\ Diehl, A.\ Donnachie, H.\,G.\ Dosch, J.\ Forshaw, 
A.\ Hebecker, P.\,V.\ Landshoff, J.\,P.\ Ma, A.\ Steinkasserer, 
and B.\,R.\ Webber for helpful discussions. 
C.\,E.\ was supported by the Bundesministerium f\"ur 
Bildung und Forschung, projects HD 05HT1VHA/0 
and HD 05HT4VHA/0, and by a Feodor Lynen fellowship of the 
Alexander von Humboldt Foundation. 

\begin{appendix}
\numberwithin{equation}{section}

\section{Propagators and decomposition of the Compton amplitude}
\label{appA}

In this appendix we give the derivation of the analytic expressions 
for the diagrams of figure \ref{fig2}. We start with the Lagrangian of QCD 
\begin{eqnarray}\label{A.1}
{\cal L}_{\mbox{\scriptsize QCD}}(x)=
-\frac{1}{2}\mathrm{Tr}\left(G_{\lambda\rho}(x)G^{\lambda\rho}(x)\right)
+\sum_q\bar{q}(x)(i\gamma^{\lambda}D_{\lambda}-m^{(0)}_q)q(x)\,,
\end{eqnarray}
where $G_{\lambda\rho}$ is the matrix-valued gluon field strength tensor and 
$D_{\lambda}$ the covariant derivative (for the notation see \cite{4}): 
\begin{eqnarray}\label{A.2}
G_{\lambda\rho}(x)
&=&
\left(\partial_{\lambda}G^a_{\rho}(x)-\partial_{\rho}G^a_{\lambda}(x)- 
g^{(0)}f_{abc}G^b_{\lambda}(x)G^c_{\rho}(x)\right)
\frac{\lambda_a}{2}\,,\\
D_{\lambda}q(x)&=&\left(\partial_{\lambda}
+ig^{(0)}G^a_{\lambda}\frac{\lambda_a}{2}\right) q(x)\,.
\label{A.3}
\end{eqnarray}
Here $g^{(0)}$ is the bare coupling parameter and $m^{(0)}_q$ are 
the bare quark masses. Since ${\cal L}_{\mbox{\scriptsize QCD}}$ 
is bilinear in the quark 
fields we can immediately perform the $q$ and $\bar{q}$ functional 
integrals in (\ref{2.6}). To write the result in a concise form 
we use the quark propagator $S^{(q)}_F(x,y;G)$ 
for given gluon potential, see (\ref{A.4}). 
This propagator satisfies Lippmann-Schwinger relations. 
Let $S_F^{(q,0)}$ be the free propagator for mass $m_q$ 
(see (\ref{2.11}), (\ref{2.12})). Then we have in matrix notation 
\bea
\label{A.4a}
S_F^{(q)} (G) &=& S_F^{(q,0)} 
- S_F^{(q,0)} 
\left(g^{(0)} \slash{G}^a \frac{\lambda_a}{2} 
+m_q^{(0)} - m_q \right) S_F^{(q)} (G) 
\nn \\
&=& S_F^{(q,0)} - S_F^{(q)} (G) \left( g^{(0)} \slash{G}^a \frac{\lambda_a}{2} 
+m_q^{(0)} - m_q \right) S_F^{(q,0)} \,.
\eea
From (\ref{A.4a}) we get 
\bea
\label{A.4b}
S_F^{(q)} (x,y;G) \left( i \leftDslash_y + m_q^{(0)} \right) 
&=& S_F^{(q)} (x,y;G) \left( i \leftslash_y + g^{(0)} \slash{G}^a 
  \frac{\lambda_a}{2} + m_q^{(0)} \right) 
\nn \\
&=& \delta^{(4)} (x-y) \,.
\eea
We define the contraction of two quark fields of flavour $q$ as 
\begin{equation}\label{A.5}
\wick{1}{<1 q(x)>1{\overline{q}}(y)=\frac{1}{i}S^{(q)}_F(x,y;G)}
\end{equation}
\newpage
\noindent
and represent it by a line in our diagrams: 
\vspace{.3cm}
\begin{center}
\includegraphics[width=3.5cm]{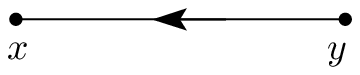}
\end{center}
In a functional integral over gluon and quark degrees of freedom we 
then have to contract all quark fields in all possible ways and integrate 
over the remaining gluon degrees of freedom including the fermion 
determinant. Schematically we have:
\begin{eqnarray}\label{A.6}
\Big\langle\ldots q(x)\bar{q}(y)\ldots\Big\rangle_{G,q,\bar{q}}&=&\Big\langle\ldots
\frac{1}{i}S^{(q)}_F(x,y;G)\ldots\Big\rangle_G \nonumber\\
&&+ \,\textup{all other contractions of quark fields.}
\end{eqnarray}
Here the functional integral on the l.h.s.\ is defined in (\ref{2.7}), 
(\ref{2.7a}), that on the r.h.s.\ in (\ref{2.9}), (\ref{2.10}).
For the functional integral in (\ref{2.6}) (there written already 
for $x'=0$) we find in this way
\begin{eqnarray}\label{A.7}
\lefteqn{\Big\langle\psi_p(y^{\prime})J^{\mu}(x')J^{\nu}(x)\overline{\psi}_p(y)
\Big\rangle_{G,q,\bar{q}}=}\nonumber\\
&=&\sum_{q^{\prime},q}Q_{q^{\prime}}Q_q\Big\langle\Gamma_{\alpha^{\prime}\beta^{\prime}
\gamma^{\prime}}
u_{\alpha^{\prime}}(y^{\prime})u_{\beta^{\prime}}(y^{\prime})d_{\gamma^{\prime}}(y^{\prime})
\bar{q}^{\prime}(x')\gamma^\mu q^{\prime}(x')\bar{q}(x)\gamma^\nu q(x)\nonumber\\
&& \hspace*{2cm}
\bar{\Gamma}_{\alpha\beta\gamma}\bar{d}_{\gamma}(y)\bar{u}_{\beta}(y)\bar{u}_{\alpha}(y)
\Big\rangle_{G,q,\bar{q}}
\nn \\
&=&\mathcal{J}^{(a)}+\ldots+\mathcal{J}^{(g)}\,.
\end{eqnarray}
The integration over quark and antiquark fields gives rise 
to a number of terms in which these fields are contracted in all 
possible ways. The resulting propagators can be understood as 
defining a quark line skeleton for the graphical representation 
of the respective term. We have classified the 
different types of diagrams in figure \ref{fig2}, and 
the terms $\mathcal{J}^{(a)}$ to $\mathcal{J}^{(g)}$ in the 
last line above are defined as corresponding to this 
classification. 

Inserting the expression (\ref{A.7}) 
in (\ref{2.6}) leads to the decomposition of the amplitude 
in (\ref{2.7a1}) with 
\begin{eqnarray}\label{A.7a}
\mathcal{M}^{(a)\mu\nu}_{s^{\prime}s}(p^{\prime},p,q)&=&
- \frac{i}{2\pi m_pZ_p}\int d^4y^{\prime}~d^4y \,
e^{ip^{\prime}y^{\prime}}
\bar{u}_{s^{\prime}}(p^{\prime})(-i\rightslash_{y^{\prime}}
+m_p) 
\nn \\
&& \int d^4x \, e^{-iqx} \left. {\cal J}^{(a)} \right|_{x'=0} 
( i\leftslash_y + m_p ) u_s(p) e^{-ipy} \,,
\end{eqnarray}
and ${\cal M}^{(b)}$ to ${\cal M}^{(g)}$ are related to 
$\mathcal{J}^{(a)},\ldots,\mathcal{J}^{(g)}$ in a completely 
analogous way. 
The expressions $\mathcal{J}^{(a)},\ldots,\mathcal{J}^{(g)}$ for $x'=0$ are 
represented in diagrammatic language in figure \ref{fig2}. 
The general rule is that for each quark line in a diagram in figure 
\ref{fig2} one has to insert a factor $-i S^{(q)}_F(x,y;G)$ according to 
(\ref{A.5}). 
The vertices with the photon just represent the coupling 
$\gamma^{\nu}Q_q$. The functional integral $\langle~\rangle_G$ 
as defined in (\ref{2.9}) has to be performed over the integrand 
resulting in this way. This  is indicated by the shading in figures \ref{fig2}a 
to \ref{fig2}g. The explicit expressions are
\begin{equation}\label{A.9}
\mathcal{J}^{(a)}=\sum_qQ^2_q\Bigg\langle\wick{2}{<1\psi_p(y^{\prime})
>1{\overline{\psi}}_p(y)}
(-1)\mathrm{Tr}\left[\gamma^{\mu}\frac{1}{i}S_F^{(q)}(x',x;G)\gamma^{\nu}
\frac{1}{i}S_F^{(q)}(x,x';G)\right]\Bigg\rangle_G \,,
\end{equation}
where
\begin{eqnarray}\label{A.10}
\wick{2}{<1\psi_p(y^{\prime})>1{\overline{\psi}}_p(y)}
&=&
\Gamma_{\alpha^{\prime}\beta^{\prime}\gamma^{\prime}}
\bar{\Gamma}_{\alpha\beta\gamma} \,
\frac{1}{i}S^{(d)}_{F\gamma^{\prime}\gamma}(y^{\prime},y;G)
\nn \\
&&{}
\biggl\{\frac{1}{i}S^{(u)}_{F\beta^{\prime}\beta}(y^{\prime},y;G)
\frac{1}{i}S^{(u)}_{F\alpha^{\prime}\alpha}(y^{\prime},y;G)-
(\alpha\leftrightarrow \beta)\biggr\}
\,,
\end{eqnarray}
further 
\be
\label{A.11}
\mathcal{J}^{(b)}=\sum_{q^{\prime},q}Q_{q^{\prime}}Q_q
\Bigg\langle\wick{2}{<1\psi_p(y^{\prime})>1{\overline{\psi}}_p(y)}
\mathrm{Tr}\left[\gamma^{\mu}\frac{1}{i}S_F^{(q^{\prime})}(x',x';G)\right]
\mathrm{Tr}\left[\gamma^{\nu}\frac{1}{i}S^{(q)}_F(x,x;G)\right]
\Bigg\rangle_G\,,
\ee
and 
\begin{eqnarray}\label{A.12}
\lefteqn{
\mathcal{J}^{(c)}=\Gamma_{\alpha^{\prime}\beta^{\prime}\gamma^{\prime}}
\bar{\Gamma}_{\alpha\beta\gamma}\Bigg\langle Q^2_d\left[\frac{1}{i}
S^{(d)}_F(y^{\prime},x';G)\gamma^{\mu}
\frac{1}{i} S^{(d)}_F(x',x;G)\gamma^{\nu}\frac{1}{i}S^{(d)}_F(x,y;G)
\right]_{\gamma'\gamma}
}
\nonumber\\
&&{}
\hspace{.3cm}
\left[\frac{1}{i}S^{(u)}_{F\beta^{\prime}\beta}(y^{\prime},y;G)
\frac{1}{i}S^{(u)}_{F\alpha^{\prime}\alpha}(y^{\prime},y;G)-
(\alpha\leftrightarrow \beta)\right]\nonumber\\
&&{}+Q^2_u\frac{1}{i}S^{(d)}_{F\gamma^{\prime}\gamma}(y^{\prime},y;G)
\nn \\
&&{}
\hspace{.3cm}
\biggl\{\left[\frac{1}{i}S^{(u)}_F(y^{\prime},x';G)\gamma^{\mu}\frac{1}{i}S^{(u)}_F(x',x;G)
\gamma^{\nu}\frac{1}{i}S^{(u)}_F(x,y;G)\right]_{\beta^{\prime}\beta}
\frac{1}{i}S^{(u)}_{F\alpha^{\prime}\alpha}(y^{\prime},y;G)
\nonumber\\
&&{}
\hspace{.3cm}
-(\alpha^{\prime}\leftrightarrow\beta^{\prime})-(\alpha\leftrightarrow\beta)
+(\alpha\leftrightarrow\beta,\alpha^{\prime}\leftrightarrow\beta^{\prime})\biggr
\}
\Bigg\rangle_G\,.
\end{eqnarray}
${\cal J}^{(d)}$ is as ${\cal J}^{(c)}$ but with $\mu \leftrightarrow \nu$ 
and $x \leftrightarrow x'$. Explicitly, 
\begin{eqnarray}\label{A.Jd}
\lefteqn{
\mathcal{J}^{(d)}=\Gamma_{\alpha^{\prime}\beta^{\prime}\gamma^{\prime}}
\bar{\Gamma}_{\alpha\beta\gamma}\Bigg\langle Q^2_d 
\left[\frac{1}{i} S^{(d)}_F(y^{\prime},x;G)\gamma^{\nu}
\frac{1}{i} S^{(d)}_F(x,x';G)\gamma^{\mu}\frac{1}{i}S^{(d)}_F(x',y;G)
\right]_{\gamma'\gamma}
}
\nonumber\\
&&{}
\hspace{.3cm}
\left[\frac{1}{i}S^{(u)}_{F\beta^{\prime}\beta}(y^{\prime},y;G)
\frac{1}{i}S^{(u)}_{F\alpha^{\prime}\alpha}(y^{\prime},y;G)-
(\alpha\leftrightarrow \beta)\right]\nonumber\\
&&{}+Q^2_u \frac{1}{i}S^{(d)}_{F\gamma^{\prime}\gamma}(y^{\prime},y;G)
\nn \\
&&{}
\hspace{.3cm}
\biggl\{\left[\frac{1}{i}S^{(u)}_F(y^{\prime},x;G)\gamma^{\nu}\frac{1}{i}S^{(u)}_F(x,x';G)
\gamma^{\mu}\frac{1}{i}S^{(u)}_F(x',y;G)\right]_{\beta^{\prime}\beta}
\frac{1}{i}S^{(u)}_{F\alpha^{\prime}\alpha}(y^{\prime},y;G)
\nonumber\\
&&{}
\hspace{.3cm}
-(\alpha^{\prime}\leftrightarrow\beta^{\prime})-(\alpha\leftrightarrow\beta)
+(\alpha\leftrightarrow\beta,\alpha^{\prime}\leftrightarrow\beta^{\prime})\biggr
\}
\Bigg\rangle_G\,.
\end{eqnarray}
We further have 
\begin{eqnarray}\label{AJe}
\lefteqn{
\mathcal{J}^{(e)} = 
\Gamma_{\alpha'\beta'\gamma'} \bar{\Gamma}_{\alpha\beta\gamma}
\Bigg\langle \bigg{\{}
Q^2_u\frac{1}{i}S^{(d)}_{F\gamma'\gamma}(y',y;G)
}
\nn 
\\
&&{}
\bigg[
\left(\frac{1}{i}S^{(u)}_F(y',x;G)\gamma^{\nu}
\frac{1}{i} S^{(u)}_F(x,y;G)\right)_{\alpha' \alpha} 
\left(\frac{1}{i}S^{(u)}_F(y',x';G)\gamma^{\mu}
\frac{1}{i} S^{(u)}_F(x',y;G)\right)_{\beta' \beta} 
\nn
\\
&&{}- (\alpha' \leftrightarrow \beta') - (\alpha \leftrightarrow  \beta) 
+ (\alpha' \leftrightarrow \beta', \alpha \leftrightarrow  \beta) 
\bigg]
\nn \\
&&{}
+ Q_u Q_d 
\bigg[
\left(\frac{1}{i}S^{(d)}_F(y',x';G)\gamma^{\mu}
\frac{1}{i} S^{(d)}_F(x',y;G)\right)_{\gamma' \gamma} 
\nn \\
&&{}
\hspace*{1.65cm}
\left(\frac{1}{i} S^{(u)}_F(y',x;G) \gamma^\nu 
\frac{1}{i} S^{(u)}_F(x,y;G) \right)_{\alpha' \alpha}
\frac{1}{i} S^{(u)}_{F\,\beta' \beta}(y',y;G) 
\nn \\
&&{}
\hspace*{1.65cm}
- (\alpha' \leftrightarrow \beta') - (\alpha \leftrightarrow  \beta) 
+ (\alpha' \leftrightarrow \beta', \alpha \leftrightarrow  \beta) 
\bigg] 
\nn \\
&&{}
+ Q_u Q_d 
\bigg[ \left( \frac{1}{i} S^{(d)}_F(y',x;G) \gamma^\nu 
\frac{1}{i} S^{(d)}_F(x,y;G) \right)_{\gamma' \gamma}
\nn\\
&&{}
\hspace*{1.65cm}
\left( \frac{1}{i} S^{(u)}_F(y',x';G) \gamma^\mu 
\frac{1}{i} S^{(u)}_F(x',y;G) \right)_{\alpha' \alpha}
\frac{1}{i} S^{(u)}_{F\,\beta' \beta}(y',y;G)
\nn\\
&&{}
\hspace*{1.65cm}
- (\alpha' \leftrightarrow \beta') - (\alpha \leftrightarrow  \beta) 
+ (\alpha' \leftrightarrow \beta', \alpha \leftrightarrow  \beta) \bigg]
\biggr\}\Bigg\rangle_G\,,
\end{eqnarray}
\begin{eqnarray}\label{AJf}
\lefteqn{
\mathcal{J}^{(f)} = 
\Gamma_{\alpha^{\prime}\beta^{\prime}\gamma^{\prime}}
\bar{\Gamma}_{\alpha\beta\gamma}
\Bigg\langle \bigg[
\sum_{q'} Q_{q'} (-1) 
\mathrm{Tr}\left(\gamma^\mu\frac{1}{i}S_F^{(q^{\prime})}(x',x';G)\right) \bigg]
}
\nn \\
&&{}\bigg{\{} Q_u \frac{1}{i}S^{(d)}_{F\gamma'\gamma}(y',y;G)
\bigg[
\left(\frac{1}{i}S^{(u)}_F(y',x;G)\gamma^{\nu} 
\frac{1}{i}S^{(u)}_F(x,y;G)\right)_{\alpha' \alpha}
\frac{1}{i}S^{(u)}_{F\beta'\beta}(y',y;G)
\nn \\
&&{}- (\alpha' \leftrightarrow \beta') - (\alpha \leftrightarrow  \beta) 
+ (\alpha' \leftrightarrow \beta', \alpha \leftrightarrow  \beta) 
\bigg]
\nn \\
&&{} +Q_d \bigg(
\frac{1}{i}S^{(d)}_F(y',x;G) \gamma^\nu
\frac{1}{i}S^{(d)}_F(x,y;G)\bigg)_{\gamma'\gamma}
\nn \\
&&{} \bigg[
\frac{1}{i}S^{(u)}_{F\alpha'\alpha}(y',y;G) 
\frac{1}{i}S^{(u)}_{F\beta' \beta}(y',y;G) 
- (\alpha \leftrightarrow  \beta) \bigg]
\bigg{\}} \Bigg\rangle_G \,,
\end{eqnarray}
\begin{eqnarray}\label{AJg}
\lefteqn{
\mathcal{J}^{(g)} = 
\Gamma_{\alpha^{\prime}\beta^{\prime}\gamma^{\prime}}
\bar{\Gamma}_{\alpha\beta\gamma}
\Bigg\langle \bigg[
\sum_{q} Q_{q} (-1) 
\mathrm{Tr}\left(\gamma^\nu\frac{1}{i}S_F^{(q)}(x,x;G)\right) \bigg]
}
\nn \\
&&{}
\bigg{\{} Q_u \frac{1}{i}S^{(d)}_{F\gamma'\gamma}(y',y;G)
\bigg[
\left(\frac{1}{i}S^{(u)}_F(y',x';G)\gamma^{\mu} 
\frac{1}{i}S^{(u)}_F(x',y;G)\right)_{\alpha' \alpha}
\frac{1}{i}S^{(u)}_{F\beta'\beta}(y',y;G)
\nn \\
&&{}- (\alpha' \leftrightarrow \beta') - (\alpha \leftrightarrow  \beta) 
+ (\alpha' \leftrightarrow \beta', \alpha \leftrightarrow  \beta) 
\bigg]
\nn \\
&&{}+Q_d \bigg(
\frac{1}{i}S^{(d)}_F(y',x';G) \gamma^\mu
\frac{1}{i}S^{(d)}_F(x',y;G)\bigg)_{\gamma'\gamma}
\nn \\
&&{} \bigg[
\frac{1}{i}S^{(u)}_{F\alpha'\alpha}(y',y;G) 
\frac{1}{i}S^{(u)}_{F\beta' \beta}(y',y;G) 
- (\alpha \leftrightarrow  \beta) \bigg]
\bigg{\}} \Bigg\rangle_G \,.
\end{eqnarray}

Finally, we indicate how we can make an analysis of the amplitude 
(\ref{2.6}) treating incoming and outgoing photons in a symmetric way. 
Using translational invariance of the functional integral we can write 
(\ref{2.6}) also as follows 
\bea
\lefteqn{
\mathcal{M}^{\mu\nu}_{s^{\prime}s}(p^{\prime},p,q)
= -\frac{i}{2 \pi m_p Z_p} \int d^4y \, d^4x' \, d^4x 
\bigg[ \bar{u}_{s^{\prime}}(p^{\prime})
(-i\rightslash_{y'}+m_p) 
}
\nn \\
&&{} 
\left\langle\psi_p(y^{\prime}) e^{i q' x'} J^{\mu}(x') 
e^{-iqx} J^\nu(x) \overline{\psi}_p(y)\right\rangle_{G,q,\bar{q}} 
(i\leftslash_y+m_p)u_s(p)e^{-ipy} \left. \bigg] 
\right|_{y'=0}  \,. 
\,\,\,\,
\eea
Inserting here the decomposition (\ref{A.7}) of the functional integral 
we obtain 
\bea
\mathcal{M}^{(a)\mu\nu}_{s's}(p',p,q)
&=& -\frac{i}{2 \pi m_p Z_p} \int d^4y \, d^4x' \, d^4x \,
e^{i q' x'} e^{-iqx}  e^{-ipy} 
\nn \\
&&{} 
\bar{u}_{s^{\prime}}(p^{\prime})(-i\rightslash_{y'}+m_p) 
\left. 
\mathcal{J}^{(a)} (i\leftslash_y+m_p)u_s(p) 
\right|_{y'=0} 
\,,
\eea
and 
similarly for $\mathcal{M}^{(b)},\dots, \mathcal{M}^{(g)}$. 
With (\ref{A.9}) we now find 
\bea
\label{A.18}
\mathcal{M}^{(a)\mu\nu}_{s's}(p',p,q)
&=& -\frac{i}{2 \pi m_p Z_p} \int d^4y  
\bigg[
\bar{u}_{s^{\prime}}(p^{\prime})(-i\rightslash_{y'}+m_p) 
\\
&&{} 
\sum_q Q_q^2 \Big\langle
\wick{2}{<1\psi_p(y^{\prime})>1{\overline{\psi}}_p(y)} 
\tilde{A}^{(q)\mu\nu}(q',q) \Big\rangle_G 
(i\leftslash_y+m_p)u_s(p)e^{-ipy} \left. \bigg] 
\right|_{y'=0} 
\,,
\nn
\eea
where 
\bea
\label{A.19}
\tilde{A}^{(q)\mu\nu}(q',q)
&=& \int d^4x'\,d^4x \,
\mathrm{Tr}\left[e^{iq'x'} \gamma^{\mu}S^{(q)}_F(x',x;G)
e^{-iqx}\gamma^{\nu}S^{(q)}_F(x,x';G) \right] 
\,. \hspace*{.3cm}
\eea
The further analysis of $\tilde{A}^{(q)\mu\nu}$ can then be done 
exactly as for $A^{(q)\mu\nu}$ of (\ref{2.8a}) in section 
\ref{sec:currenttoquark}. The representation (\ref{A.18}), (\ref{A.19}) 
for the amplitude then also allows the treatment of the 
final state photon $\gamma(q',\mu)$ in the same way as the 
initial state photon, that is replacing it by a dipole in 
the way explained in section \ref{nonpertsect}. 

\section{Decomposition using spin sums}
\label{appB}

The free quark propagator for mass $m_q$ can be written as follows
\begin{eqnarray}\label{B.1}
S^{(q,0)}_F(x,y)&=&-\int\frac{d^4k}{(2\pi)^4}\,e^{-ik (x-y)}
\frac{\kslash+m_q}{k^2-m^2_q+i\epsilon}
\, \mathbf{1}_3
\nonumber\\
&=&\theta(x^0-y^0)\frac{i}{(2\pi)^3}\int\frac{d^3k}{2k^0}\,
e^{-ik(x-y)}(\kslash+m_q)
\, \mathbf{1}_3
\nonumber\\
&&{}+\theta(y^0-x^0)\frac{i}{(2\pi)^3}\int\frac{d^3k}{2k^0}\,
e^{ik(x-y)}(-\kslash+m_q) 
\, \mathbf{1}_3
\,,
\end{eqnarray}
where $k^0=(\mathbf{k}^2+m^2_q)^{1/2}$, 
and $\mathbf{1}_3$ is the unit matrix in colour space. 
We define the Dirac times colour spinors as 
\bea
\label{colorspinoru}
u_r(k) &=& u_\lambda(k) \,\mathbf{e}_A \,, \\
\label{colorspinorv}
v_r(k) &=& v_\lambda(k) \,\mathbf{e}_A \,. 
\eea
Here $r=(\lambda,A)$ are the spin times colour 
indices with $\lambda \in \{ 1/2, -1/2\}$ and $A \in \{ 1,2,3\}$. 
$u_\lambda(k)$ and $v_\lambda(k)$ are the 
usual Dirac spinors for mass $m_q$, and $\mathbf{e}_A$ 
are orthonormal basis vectors in colour space. We then 
have for the spin and colour sums 
\bea
\label{spinsumu}
\sum_r u_r(k) \bar{u}_r(k) &=& (\slash{k} + m_q)
\, \mathbf{1}_3 \,,
\\
\label{spinsumv}
\sum_r v_r(k) \bar{v}_r(k) &=& (\slash{k} - m_q)
\, \mathbf{1}_3 \,.
\eea
Inserting this and the 
Fourier representations for the $\theta$-functions 
in (\ref{B.1}) we get 
\begin{eqnarray}\label{B.2}
S^{(q,0)}_F(x,y)&=&-\frac{1}{2\pi}\int^{\infty}_{-\infty}\frac{d\omega}{\omega+i\epsilon}
\int\frac{d^3k}{(2\pi)^32k^0}\nonumber\\
&&{}
\sum_r\left\{e^{-ik_\omega x}u_r(k)\bar{u}_r(k)
e^{ik_\omega y}-e^{ik_\omega x}v_r(k)\bar{v}_r(k)e^{-ik_\omega y}\right\} 
\,.
\end{eqnarray}
Here
\begin{eqnarray}\label{B.3}
k&=&\left(\begin{array}{c}k^0 \\ \mathbf{k}\end{array}\right),\nonumber\\
k_\omega&=&\left(\begin{array}{c}k^0+\omega \\ \mathbf{k} \end{array}\right)\,.
\end{eqnarray}
We find it convenient to introduce matrix notation. For this we define
\begin{eqnarray}\label{B.4}
(x|k_\omega,r)&=&u_r(k)e^{-ik_\omega x}\,,\nonumber\\
(k_\omega,r|x)&=&e^{ik_\omega x}\bar{u}_r(k)\,,\nonumber\\
(x|\overline{k_\omega,r})&=&v_r(k)e^{ik_\omega x}\,,\nonumber\\
(\overline{k_\omega,r}|x)&=&e^{-ik_\omega x}\bar{v}_r(k) \,.
\end{eqnarray}
We also use the summation convention for spin times colour indices and 
momenta, implying integration over two identical momenta $k$ with 
the usual invariant phase space measure. 
Note that this measure contains the energy 
denominator $k^0$ and not $k_\omega^0$. 
To give an example we write
\begin{eqnarray}\label{B.5}
(x|k_\omega,r)(k_\omega,r|y)=\sum_r\int
\frac{d^3 k}{(2\pi)^32k^0} \, e^{-ik_\omega x} u_r(k)\bar{u}_r(k)e^{ik_\omega y} \,.
\end{eqnarray}
In this way we can represent the free propagator (\ref{B.1}) as
\begin{eqnarray}\label{B.6}
S^{(q,0)}_F(x,y)&=&(x|S^{(q,0)}_F|y)\nonumber\\
&=&-\frac{1}{2\pi}\int^{\infty}_{-\infty}\frac{d\omega}{\omega+i\epsilon}\,
\left[(x|k_\omega,r)(k_\omega,r|y)
-(x|\overline{k_\omega,r})(\overline{k_\omega,r}|y)\right] \,.
\end{eqnarray}
In (\ref{B.6}) the free propagator $S^{(q,0)}_F$ is written as the sum of dyadic products
of quark and antiquark spinors where the energies in the plane wave factors in
(\ref{B.4}) are off the energy shell.

In the remaining part of this appendix we give the derivation of (\ref{2.14}). Integration 
by parts and insertion of (\ref{B.6}) for the free propagators leads from (\ref{2.13}) to
\begin{eqnarray}\label{B.7}
A^{(q)\mu\nu}(q)&=&\int d^4x~d^4z~d^4z^{\prime} \,
\nn \\
&&{}
\mathrm{Tr}\biggl\{
\left[
(-i\rightslash_{z'}+m_q) S^{(q)}_F(z^{\prime},0;G) 
\gamma^\mu S^{(q)}_F(0,z;G) (i\leftslash_z+m_q)
\right]
\nn \\
&&{}
S^{(q,0)}_F(z,x) e^{-iqx}\gamma^{\nu} S^{(q,0)}_F(x,z^{\prime})
\biggl\}
\\
\label{B.7a}
&=&{}
\frac{1}{(2\pi)^2}\int^{\infty}_{-\infty}\frac{d\omega}{\omega+i\epsilon}
\int^{\infty}_{-\infty}\frac{d\omega^{\prime}}{\omega^{\prime}+i\epsilon}
\int d^4x~d^4z~d^4z^{\prime}\nonumber\\
&&{}
\mathrm{Tr}\biggl\{
\left[(-i\rightslash_{z^{\prime}}+m_q)S^{(q)}_F(z^{\prime},0;G)
\gamma^{\mu}S^{(q)}_F(0,z;G)(i\leftslash_z+m_q)\right]
\nonumber\\
&&{}
\big[(z|k_\omega,r)(k_\omega,r|x)-(z|\overline{k_\omega,r})(\overline{k_\omega,r}|x)\big]
e^{-iqx}\gamma^{\nu}\nonumber\\
&&{}
\big[(x|k^{\prime}_{\omega^{\prime}},r^{\prime})
(k^{\prime}_{\omega^{\prime}},r^{\prime}|z^{\prime})
-(x|\overline{k^{\prime}_{\omega^{\prime}},r^{\prime}})(\overline{k^{\prime}_{\omega^\prime},
r^{\prime}}|z^{\prime})\big]\biggr\}\,,
\end{eqnarray}
where in analogy to (\ref{B.3}) we have defined
\begin{eqnarray}\label{B.3strich}
k'&=&\left(\begin{array}{c}k'^0 \\ \mathbf{k}'\end{array}\right),\nonumber\\
k'_{\omega'}&=&\left(\begin{array}{c}k'^0+\omega' \\ \mathbf{k}' \end{array}\right)\,.
\end{eqnarray}
We define
\be
\label{B.8}
(z^{\prime}|B^{(q) \mu}|z)=(-i\rightslash_{z^{\prime}}+m_q)S^{(q)}_F(z^{\prime},0;G)
\gamma^{\mu}S^{(q)}_F(0,z;G)(i\leftslash_z+m_q)\,,
\ee
and further 
\bea
\label{B.10}
(k^{\prime}_{\omega^{\prime}},r^{\prime}|B^{(q) \mu}|k_\omega,r)
&=&\int d^4z~d^4z^{\prime}~
(k^{\prime}_{\omega^{\prime}},r'|z^{\prime})(z^{\prime}|B^{(q) \mu}|z)(z|k_\omega,r)\,,
\\
\label{B.12}
(k^{\prime}_{\omega^{\prime}},r^{\prime}|B^{(q) \mu}|\overline{k_\omega,r})&=&
\int d^4z~d^4z^{\prime}~(k^{\prime}_{\omega^{\prime}},r^{\prime}|z^{\prime})
(z^{\prime}|B^{(q) \mu}|z)(z|\overline{k_\omega,r})\,,
\\
\label{B.9}
(\overline{k^{\prime}_{\omega^{\prime}},r^{\prime}}|B^{(q) \mu}|k_\omega,r)&=&
\int d^4z~d^4z^{\prime}~
(\overline{k^{\prime}_{\omega^{\prime}},r^{\prime}}|z^{\prime})
(z^{\prime}|B^{(q) \mu}|z)(z|k_\omega,r)\,,
\\
\label{B.11}
(\overline{k^{\prime}_{\omega^{\prime}},
r^{\prime}}|B^{(q) \mu}|\overline{k_\omega,r})&=&\int d^4z~d^4z^{\prime}~
(\overline{k^{\prime}_{\omega^{\prime}},r^{\prime}}|z^{\prime})(z^{\prime}|B^{(q) \mu}|z)(z|
\overline{k_\omega,r})\,,
\eea
and 
\begin{eqnarray}\label{B.14}
(k_\omega,r|E^{\nu}(q)|k^{\prime}_{\omega^{\prime}},r^{\prime})&=&\int d^4x~
(k_\omega,r|x)e^{-iqx}\gamma^{\nu}(x|k^{\prime}_{\omega^{\prime}},r^{\prime})\nonumber\\
&=&(2\pi)^4\delta^{(4)}(k_\omega-k^{\prime}_{\omega^{\prime}}-q)
\bar{u}_r(k)\gamma^{\nu}u_{r'} (k^{\prime})\,,
\\
\label{B.13}
(k_\omega,r|E^{\nu}(q)|\overline{k^{\prime}_{\omega^{\prime}},r^{\prime}})
&=&\int d^4x~(k_\omega,r|x)e^{-iqx}\gamma^{\nu}
(x|\overline{k^{\prime}_{\omega^{\prime}},r^{\prime}})\nonumber\\
&=&(2\pi)^4\delta^{(4)}(k_\omega+k^{\prime}_{\omega^{\prime}}-q)
\bar{u}_r(k)\gamma^{\nu}v_{r^{\prime}}(k^{\prime})\,,
\\
\label{B.16}
(\overline{k_\omega,r}|E^{\nu}(q)|k^{\prime}_{\omega^{\prime}},r^{\prime})&=&\int d^4x~
(\overline{k_\omega,r}|x)e^{-iqx}\gamma^{\nu}(x|k^{\prime}_{\omega^{\prime}},r^{\prime})
\nonumber\\
&=&(2\pi)^4\delta^{(4)}(-k_\omega-k^{\prime}_{\omega^{\prime}}-q)\bar{v}_r(k)
\gamma^{\nu}u_{r'}(k^{\prime})\,,
\\
\label{B.15}
(\overline{k_\omega,r}|E^{\nu}(q)|\overline{k^{\prime}_{\omega^{\prime}},r^{\prime}})
&=&\int d^4x~
(\overline{k_\omega,r}|x)e^{-iqx}\gamma^{\nu}
(x|\overline{k^{\prime}_{\omega^{\prime}},r^{\prime}})\nonumber\\
&=&(2\pi)^4\delta^{(4)}(-k_\omega+k^{\prime}_{\omega^{\prime}}-q)\bar{v}_r(k)
\gamma^{\nu}v_{r^{\prime}}(k^{\prime})\,.
\end{eqnarray}
Note that the matrix elements of $E^\nu$ depend on $q$, in contrast to 
those of $B^{(q) \mu}$. We find from (\ref{B.7a})
\begin{eqnarray}\label{B.17}
A^{(q)\mu\nu}(q)
&=&\frac{1}{(2\pi)^2}\int\frac{d\omega}{\omega+i\epsilon}\int
\frac{d\omega^{\prime}}{\omega^{\prime}+i\epsilon}
\nn\\
&&{}
\biggl\{
-(\overline{k^{\prime}_{\omega^{\prime}},r^{\prime}}
|B^{(q) \mu}|k_\omega,r)(k_\omega,r|E^{\nu}(q)|
\overline{k^{\prime}_{\omega^{\prime}},r^{\prime}})
\nn \\
&&{}
\quad
+(k^{\prime}_{\omega^{\prime}},r^{\prime}|B^{(q) \mu}|k_\omega,r)
(k_\omega,r|E^{\nu}(q)|k^{\prime}_{\omega^{\prime}},r^{\prime})
\nonumber\\
&&{}
\quad-(k^{\prime}_{\omega^{\prime}},r^{\prime}|B^{(q) \mu}|\overline{k_\omega,r})
(\overline{k_\omega,r}|E^{\nu}(q)|k^{\prime}_{\omega^{\prime}},r^{\prime})
\nonumber\\
&&{}
\quad+
(\overline{k^{\prime}_{\omega^{\prime}},r^{\prime}}|B^{(q) \mu}|\overline{k_\omega,r})
(\overline{k_\omega,r}|E^{\nu}(q)|
\overline{k^{\prime}_{\omega^{\prime}},r^{\prime}})\biggr\}\,.
\end{eqnarray}
Inserting the explicit expressions (\ref{B.14}) to (\ref{B.15}) for the 
matrix elements of $E^{\nu}(q)$ and restoring all sums and integrations 
we find
\begin{eqnarray}\label{B.18}
\lefteqn{A^{(q)\mu\nu}(q)=(2\pi)^2\int\frac{d\omega}
{\omega+i\epsilon}\int\frac{d\omega^{\prime}}
{\omega^{\prime}+i\epsilon}\int\frac{d^3k}{(2\pi)^32k^0}
\int\frac{d^3k^{\prime}}{(2\pi)^32k^{\prime 0}}\sum_{r,r^{\prime}}}
\\
&&{}\hspace*{-.4cm}
\biggl\{ -\delta(k^0+k^{\prime 0}-q^0+\omega+\omega^{\prime})\delta^{(3)}
(\mathbf{k}+\mathbf{k}^{\prime}-\mathbf{q})
(\overline{k^{\prime}_{\omega^{\prime}},r^{\prime}}|B^{(q) \mu}|k_\omega,r)
\bar{u}_r(k)\gamma^{\nu}v_{r^{\prime}}(k^{\prime})
\nonumber\\
&&{}\hspace*{-.4cm}
\quad +\delta(k^0-k^{\prime 0}-q^0+\omega-\omega^{\prime})
\delta^{(3)}(\mathbf{k}-\mathbf{k}^{\prime}-\mathbf{q})
(k^{\prime}_{\omega^{\prime}},r^{\prime}|B^{(q) \mu}|k_\omega,r)
\bar{u}_r(k)\gamma^{\nu}u_{r^{\prime}}(k^{\prime})
\nonumber\\
&&{}\hspace*{-.4cm}
\quad -\delta(-k^0-k^{\prime 0}-q^0-\omega-\omega^{\prime})
\delta^{(3)}(-\mathbf{k}-\mathbf{k}^{\prime}-\mathbf{q})
(k^{\prime}_{\omega^{\prime}},r^{\prime}|B^{(q) \mu}|
\overline{k_\omega,r})\bar{v}_r(k)\gamma^{\nu}u_{r^{\prime}}(k^{\prime})
\nonumber \\
&&{}\hspace*{-.4cm}
\quad+\delta(-k^0+k^{\prime 0}-q^0-\omega+\omega^{\prime})
\delta^{(3)}(-\mathbf{k}+\mathbf{k}^{\prime}-\mathbf{q})
(\overline{k^{\prime}_{\omega^{\prime}},r^{\prime}}|B^{(q) \mu}|
\overline{k_\omega,r})\bar{v}_r(k)\gamma^{\nu}v_{r^{\prime}}(k^{\prime})\biggr\}\,.
\nonumber
\end{eqnarray}
Performing the $\omega^\prime$ and $\mathbf{k}^{\prime}$ integrations leads to
\begin{eqnarray}\label{B.19}
\lefteqn{A^{(q)\mu\nu}(q)=\frac{1}{2\pi}\int\frac{d\omega}{\omega+i\epsilon}\int
\frac{d^3k}{(2\pi)^32k^0}\sum_{r,r^{\prime}}}
\\
&&{}
\biggl\{-(\overline{k^{\prime (1)}_\omega,r^{\prime}}
|B^{(q) \mu}|k_\omega,r)\bar{u}_r(k)\gamma^{\nu}v_{r^{\prime}}(k^{\prime (1)})
(q^0-k^0-k^{\prime (1)0}-\omega+i\epsilon)^{-1}(2k^{\prime (1)0})^{-1}
\nonumber\\
&&{}
\quad+(k^{\prime(2)}_\omega,r^{\prime}|B^{(q) \mu}|k_\omega,r)
\bar{u}_r(k)\gamma^{\nu}u_{r^{\prime}}(k^{\prime(2)})
(-q^0+k^0-k^{\prime(2)0}+\omega+i\epsilon)^{-1}(2k^{\prime(2)0})^{-1}
\nonumber\\
&&{}
\quad-(k^{\prime(3)}_\omega,r^{\prime}|B^{(q) \mu}|\overline{k_\omega,r})
\bar{v}_r(k)\gamma^{\nu}u_{r^\prime}(k^{\prime(3)})
(-q^0-k^0-k^{\prime(3)0}-\omega+i\epsilon)^{-1}(2k^{\prime(3)0})^{-1}
\nonumber\\
&&{}
\quad+(\overline{k^{\prime(4)}_\omega,r^\prime}|B^{(q) \mu}|\overline{k_\omega,r})\bar{v}_r(k)
\gamma^\nu v_{r^\prime}(k^{\prime(4)})
(q^0+k^0-k^{\prime(4)0}+\omega+i\epsilon)^{-1}(2k^{\prime(4)0})^{-1}\biggr\}\,,
\nonumber 
\end{eqnarray}
where due to the delta functions in the $\omega'$ integrations 
the $k^{\prime(i)}$ satisfy relations similar to (\ref{B.3}) and (\ref{B.3strich}). 
Explicitly, 
\begin{eqnarray}
\label{B.20}
\mathbf{k}^{\prime(1)}=\mathbf{k}^{\prime(1)}_\omega 
\!\!\!&=& \!\!\!\mathbf{q}-\mathbf{k}\,,
\nonumber\\
k^{\prime(1)0}&=&\sqrt{m^2_q+(\mathbf{k}^{\prime(1)})^2}\,,
\\
k^{\prime(1)0}_\omega&=&q^0-k^0-\omega\,,
\nn
\end{eqnarray}
\begin{eqnarray}\label{B.21}
\mathbf{k}^{\prime(2)}=\mathbf{k}^{\prime(2)}_\omega
\!\!\!&=& \!\!\! -\mathbf{q}+\mathbf{k}\,,
\nonumber\\
k^{\prime(2)0}&=&\sqrt{m^2_q+(\mathbf{k}^{\prime(2)})^2}\,,
\\
k^{\prime(2)0}_\omega&=&-q^0+k^0+\omega\,,
\nn
\end{eqnarray}
\begin{eqnarray}\label{B.22}
\mathbf{k}^{\prime(3)}=\mathbf{k}^{\prime(3)}_\omega
\!\!\!&=& \!\!\! -\mathbf{q}-\mathbf{k}\,,
\nonumber\\
k^{\prime(3)0}&=&\sqrt{m^2_q+(\mathbf{k}^{\prime(3)})^2}\,,
\\
k^{\prime(3)0}_\omega&=&-q^0-k^0-\omega\,,
\nn
\end{eqnarray}
\begin{eqnarray}\label{B.23}
\mathbf{k}^{\prime(4)}=\mathbf{k}^{\prime(4)}_\omega
\!\!\!&=& \!\!\!  \mathbf{q}+\mathbf{k}\,,
\nonumber\\
k^{\prime(4)0}&=&\sqrt{m^2_q+(\mathbf{k}^{\prime(4)})^2}\,,
\\
k^{\prime(4)0}_\omega&=&q^0+k^0+\omega\,.
\nn
\end{eqnarray}

\boldmath
\section{The matrix element ${\cal M}^{(b)}$}
\unboldmath
\label{appC}

In this appendix we discuss the amplitude ${\cal M}^{(b)}$ 
(\ref{Mbexpl}) in a similar way as we did for ${\cal M}^{(a)}$ 
in sections \ref{sec:currenttoquark} and \ref{sec:currenttoquarkqft}. 
Our aim is to derive (\ref{2.44}) and (\ref{2.45}). 

We start from (\ref{Mbexpl}) and write ${\cal M}^{(b)}$ as 
\bea
\label{C.1}
\mathcal {M}^{(b) \mu \nu}_{s's} (p',p,q) &=& 
- \frac{i}{2 \pi m_p Z_p} \sum_{q',q} Q_{q'} Q_q 
\nn \\
&&{}
\left\langle \left(p',s' |\Psi|p,s \right) (-1) \mathrm{Tr}
\left[ \gamma^\mu S_F^{(q')}(0,0;G) \right] A_b^{(q)\nu}(q) 
\right\rangle_G \,,
\eea
where 
\bea
\label{C.2}
\left(p',s' |\Psi|p,s \right) &=& \int d^4y'\, d^4y \,
e^{ip^{\prime}y^{\prime}}\bar{u}_{s^{\prime}}(p^{\prime})
(-i\rightslash_{y^{\prime}} +m_p) 
\nn \\
&&{}
\wick{2}{<1\psi_p(y^{\prime})>1{\overline{\psi}}_p(y)} 
(i\leftslash_y+m_p)u_s(p)e^{-ipy} 
\eea
and 
\be
\label{C.3}
 A_b^{(q)\nu}(q) = \int d^4x \, 
\mathrm{Tr} \left[ e^{-iqx} \gamma^\nu  S_F^{(q)}(x,x;G) \right] 
\,.
\ee
Inserting in (\ref{C.3}) factors of $1$ from (\ref{2.11}) and (\ref{2.12}) 
we get after integrating by parts 
\bea
\label{C.4}
A_b^{(q)\nu}(q) &=& \int d^4x \, d^4z \, d^4z' \,
\mathrm{Tr} \biggl\{
\left[(-i\rightslash_{z^{\prime}}+m_q)S^{(q)}_F(z^{\prime},z;G)
(i\leftslash_z+m_q)\right]
\nn \\
&&{}
S_F^{(q,0)}(z,x) e^{-iqx} \gamma^\nu S_F^{(q,0)}(x,z')
\biggl\}
\,.
\eea
Thus $A_b^{(q)\nu}(q)$ has the same structure as $A^{(q)\mu\nu}(q)$ 
(\ref{B.7}) with the replacement 
\be
\label{C.5}
S^{(q)}_F(z',0;G) \gamma^\mu S^{(q)}_F(0,z;G) 
\,\,\,\longrightarrow \,\,\,
S^{(q)}_F(z',z;G) 
\,.
\ee
The next step is to insert for $S_F^{(q,0)}$ in (\ref{C.4}) the expansion 
(\ref{B.6}). Then all remaining steps done for $A^{(q)\mu\nu}(q)$ 
in appendix \ref{appB} can be repeated for $A_b^{(q)\nu}(q)$. 
With the replacement (\ref{C.5}) we define the quantities analogous 
to (\ref{B.8})-(\ref{B.11}) 
\be
\label{C.6}
(z^{\prime}|C^{(q)}|z)=(-i\rightslash_{z'}+m_q)S^{(q)}_F(z',z;G)
(i\leftslash_z+m_q)\,,
\ee
\be
\label{C.7}
(k^{\prime}_{\omega^{\prime}},r^{\prime}|C^{(q)}|k_\omega,r)
=\int d^4z~d^4z'\,
(k^{\prime}_{\omega^{\prime}},r'|z^{\prime})(z^{\prime}|C^{(q)}|z)(z|k_\omega,r)\,,
\ee
etc. In analogy to (\ref{2.14})-(\ref{2.17b}) we get then 
\be
\label{C.8}
A_b^{(q)\nu}(q) = \sum^4_{j=1}A^{(q)\nu}_{b,j}(q)\,,
\ee
where 
\be
\label{C.9}
A^{(q)\nu}_{b,j}(q) = \frac{1}{2\pi} \int \frac{d\omega}{\omega+i\epsilon}
\int \frac{d^3k}{(2\pi)^3 2k^0} \, b_j\,. 
\ee
The $b_j$ ($j=1,\dots,4$) are defined as the $a_j$ 
in (\ref{2.16})-(\ref{2.17b}) with the replacement 
\be
\label{C.10}
B^{(q)\mu} \,\,\, \longrightarrow \,\,\, C^{(q)} \,.
\ee
Inserting now (\ref{C.8}) in (\ref{C.1}) we get 
\be
\label{C.11}
\mathcal {M}^{(b) \mu \nu}_{s's} (p',p,q) =
\sum_{j=1}^4 \mathcal {M}^{(b,j) \mu \nu}_{s's} (p',p,q) 
\,,
\ee
where 
\bea
\label{C.12}
\mathcal {M}^{(b,j) \mu \nu}_{s's} (p',p,q) &=& 
-\frac{i}{2 \pi m_p Z_p} \sum_{q',q} Q_{q'} Q_q 
\\
&&{}
\left\langle \left(p',s' |\Psi|p,s \right) (-1) \mathrm{Tr}
\left[ \gamma^\mu S_F^{(q')}(0,0;G) \right] A_{b,j}^{(q)\nu}(q) 
\right\rangle_G .
\nn
\eea
Explicitly we get with the momenta $k'^{(j)}$ and $k'^{(j)}_\omega$ 
as in (\ref{B.20})-(\ref{B.23}) 
\bea
\label{C.13}
\mathcal {M}^{(b,1) \mu \nu}_{s's} (p',p,q) 
\!&=&\! 
\frac{1}{2\pi} \sum_{q',q} Q_{q'} Q_q 
\int\frac{d\omega}{\omega+i\epsilon}
\int\frac{d^3k}{(2\pi)^3 2k^0}
(q^0-k^0-k^{{\prime}(1)0}-\omega+i\epsilon)^{-1}
\nn \\
&&{}
(2k^{{\prime}(1)0})^{-1}
\sum_{r^{\prime},r}
\langle \gamma(q^{\prime},\mu), p(p^{\prime},s^{\prime})
|\mathcal{T}^{(q',b)}|\bar{q}(k^{{\prime}(1)}_\omega,r^{\prime}), 
q(k_\omega,r),p(p,s)\rangle\nonumber\\
&&{}
\bar{u}_r(k)Z_q\gamma^{\nu}v_{r^{\prime}}(k^{{\prime}(1)})\,,
\eea
\bea
\label{C.14}
\mathcal{M}^{(b,2)\mu\nu}_{s^{\prime}s}(p^{\prime},p,q)
\!&=&\!
\frac{1}{2\pi} \sum_{q',q} Q_{q'} Q_q 
\int\frac{d\omega}{\omega+i\epsilon}
\int\frac{d^3k}{(2\pi)^32k^0}
(-q^0+k^0-k^{{\prime}(2)0}+\omega+i\epsilon)^{-1}
\nonumber\\
&&{}
(2k^{{\prime}(2)0})^{-1}
\sum_{r^{\prime},r}\langle 
\gamma(q^{\prime},\mu),q(k^{{\prime}(2)}_\omega,r^{\prime}),
p(p^{\prime},s^{\prime})|\mathcal{T}^{(q',b)}|q(k_\omega,r),p(p,s)
\rangle\nonumber\\
&&{}
\bar{u}_r(k)Z_q\gamma^{\nu}u_{r^{\prime}}(k^{{\prime}(2)})\,,
\eea
\bea
\label{C.15}
\mathcal{M}^{(b,3)\mu\nu}_{s^{\prime}s}(p^{\prime},p,q)
\!&=&\!
\frac{1}{2\pi} \sum_{q',q} Q_{q'} Q_q 
\int\frac{d\omega}{\omega+i\epsilon}
\int\frac{d^3k}{(2\pi)^32k^0}
(-q^0-k^0-k^{{\prime}(3)0}-\omega+i\epsilon)^{-1}
\nonumber\\
&&{}
(2k^{{\prime}(3)0})^{-1}
\sum_{r^{\prime},r}\langle 
\gamma(q^{\prime},\mu), 
\bar{q}(k_\omega,r), q(k^{{\prime}(3)}_\omega,r^{\prime}),
p(p^{\prime},s^{\prime})|\mathcal{T}^{(q',b)}|p(p,s)\rangle\nonumber\\
&&{}
\bar{v}_r(k)Z_q\gamma^{\nu}u_{r^{\prime}}(k^{{\prime}(3)})\,,
\eea
\bea
\label{C.16}
\mathcal{M}^{(b,4)\mu\nu}_{s^{\prime}s}(p^{\prime},p,q)
\!&=&\!
\frac{1}{2\pi} \sum_{q',q} Q_{q'} Q_q 
\int\frac{d\omega}{\omega+i\epsilon}
\int\frac{d^3k}{(2\pi)^32k^0}
(q^0+k^0-k^{{\prime}(4)0}+\omega+i\epsilon)^{-1}
\nonumber\\
&&{}
(2k^{{\prime}(4)0})^{-1}
\sum_{r^{\prime},r}\langle 
\gamma(q^{\prime},\mu),
\bar{q}(k_\omega,r), 
p(p^{\prime},s^{\prime})|\mathcal{T}^{(q',b)}|\bar{q}(k^{{\prime}(4)}_\omega,
r^{\prime}),p(p,s)\rangle\nonumber\\
&&{}
(-1) \bar{v}_r(k)Z_q\gamma^{\nu}v_{r^{\prime}}(k^{{\prime}(4)})\,, 
\eea
where 
\bea
\label{C.17}
&&{}
\hspace*{-.5cm}
\langle \gamma(q^{\prime},\mu),p(p^{\prime},s^{\prime})|\mathcal{T}^{(q',b)}|
\bar{q}(k^{{\prime}(1)}_\omega,r^{\prime}),q(k_\omega,r),p(p,s)\rangle =
\\
&&{}
\hspace*{.5cm}
= -\frac{i}{2\pi m_p Z_pZ_q} 
\left\langle
\left(p',s' |\Psi|p,s \right) 
\mathrm{Tr} \left[ \gamma^\mu S_F^{(q')}(0,0;G) \right] 
(\overline{k_\omega^{{\prime}(1)},r^{\prime}}|C^{(q)}|k_\omega,r)
\right\rangle_G ,
\nn 
\\
\label{C.18}
&&{}
\hspace*{-.5cm}
\langle \gamma(q^{\prime},\mu), 
q(k^{{\prime}(2)}_\omega,r^{\prime}),
p(p^{\prime},s^{\prime})|\mathcal{T}^{(q',b)}|q(k_\omega,r),p(p,s)\rangle = 
\\
&&{}
\hspace*{.5cm}
= -\frac{i}{2\pi m_p Z_pZ_q} 
\left\langle
\left(p',s' |\Psi|p,s \right) (-1)
\mathrm{Tr} \left[ \gamma^\mu S_F^{(q')}(0,0;G) \right] 
(k^{{\prime}(2)}_\omega,r^{\prime}|C^{(q)}|k_\omega,r)
\right\rangle_G ,
\nn
\\
\label{C.19}
&&{}
\hspace*{-.5cm}
\langle \gamma(q^{\prime},\mu),\bar{q}(k_\omega,r),
q(k^{{\prime}(3)}_\omega,r^{\prime}),
p(p^{\prime},s^{\prime})|\mathcal{T}^{(q',b)}|p(p,s)\rangle =
\\
&&{}
\hspace*{.5cm}
=-\frac{i}{2\pi m_p Z_pZ_q} 
\left\langle
\left(p',s' |\Psi|p,s \right) 
\mathrm{Tr} \left[ \gamma^\mu S_F^{(q')}(0,0;G) \right] 
(k^{{\prime}(3)}_\omega,r^{\prime}|C^{(q)}|\overline{k_\omega,r})
\right\rangle_G ,
\nn
\\
\label{C.20}
&&{}
\hspace*{-.5cm}
\langle \gamma(q^{\prime},\mu),
\bar{q}(k_\omega,r),p(p^{\prime},s^{\prime})|\mathcal{T}^{(q',b)}|\bar{q}
(k^{{\prime}(4)}_\omega,r^{\prime}),p(p,s)\rangle =
\\
&&{}
\hspace*{.5cm}
=-\frac{i}{2\pi m_p Z_pZ_q} 
\left\langle
\left(p',s' |\Psi|p,s \right) 
\mathrm{Tr} \left[ \gamma^\mu S_F^{(q')}(0,0;G) \right] 
(\overline{k^{{\prime}(4)}_\omega,r^{\prime}}|C^{(q)}|
\overline{k_\omega,r})
\right\rangle_G .
\nn
\eea
In (\ref{C.13})-(\ref{C.16}) we can replace $Z_q\gamma^\nu$ 
by (\ref{2.31}). Choosing the momenta in (\ref{C.13}) 
according to  (\ref{2.32}) leads to (\ref{2.45}). 

\end{appendix}


\begin{thebibliography}{99}

\bibitem{Nikolaev:1990ja}
N.~N.~Nikolaev and B.~G.~Zakharov,
Z.\ Phys.\ C {\bf 49} (1991) 607.

\bibitem{Nikolaev:et}
N.~N.~Nikolaev and B.~G.~Zakharov,
Z.\ Phys.\ C {\bf 53} (1992) 331.

\bibitem{Mueller:1993rr}
A.~H.~Mueller,
Nucl.\ Phys.\ B {\bf 415} (1994) 373.

\bibitem{Mueller:gb}
A.~H.~Mueller,
Nucl.\ Phys.\ B {\bf 437} (1995) 107
[arXiv:hep-ph/9408245].

\bibitem{Gribov:1968gs}
V.~N.~Gribov,
Sov.\ Phys.\ JETP {\bf 30} (1970) 709
[Zh.\ Eksp.\ Teor.\ Fiz.\  {\bf 57} (1969) 1306].

\bibitem{Ioffe:1969kf}
B.~L.~Ioffe,
Phys.\ Lett.\ B {\bf 30} (1969) 123.

\bibitem{Donnachie:en}
A.~Donnachie, H.\,G.~Dosch, O.~Nachtmann and P.\,V.~Landshoff,
{\sl Pomeron Physics And QCD}, Cambridge University Press 2002

\bibitem{Golec-Biernat:1998js}
K.~Golec-Biernat and M.~W\"usthoff,
Phys.\ Rev.\ D {\bf 59} (1999) 014017
[arXiv:hep-ph/9807513].

\bibitem{Golec-Biernat:1999qd}
K.~Golec-Biernat and M.~W\"usthoff,
Phys.\ Rev.\ D {\bf 60} (1999) 114023
[arXiv:hep-ph/9903358].

\bibitem{Golec-Biernat:2001mm}
K.~Golec-Biernat and M.~W\"usthoff,
Eur.\ Phys.\ J.\ C {\bf 20} (2001) 313
[arXiv:hep-ph/0102093].

\bibitem{Forshaw:1999uf}
J.~R.~Forshaw, G.~Kerley and G.~Shaw,
Phys.\ Rev.\ D {\bf 60} (1999) 074012
[arXiv:hep-ph/9903341].

\bibitem{Forshaw:1999ny}
J.~R.~Forshaw, G.~R.~Kerley and G.~Shaw,
Nucl.\ Phys.\ A {\bf 675} (2000) 80
[arXiv:hep-ph/9910251].

\bibitem{Forshaw:2003ki}
J.~R.~Forshaw, R.~Sandapen and G.~Shaw,
Phys.\ Rev.\ D {\bf 69} (2004) 094013
[arXiv:hep-ph/0312172].

\bibitem{Forshaw:2006np}
J.~R.~Forshaw, R.~Sandapen and G.~Shaw,
arXiv:hep-ph/0608161.

\bibitem{Donnachie:2001wt}
A.~Donnachie and H.~G.~Dosch,
Phys.\ Rev.\ D {\bf 65} (2002) 014019
[arXiv:hep-ph/0106169].

\bibitem{Dosch:1997nw}
H.~G.~Dosch, T.~Gousset and H.~J.~Pirner,
Phys.\ Rev.\ D {\bf 57} (1998) 1666
[arXiv:hep-ph/9707264].

\bibitem{D'Alesio:1998sf}
U.~D'Alesio, A.~Metz and H.~J.~Pirner,
Eur.\ Phys.\ J.\ C {\bf 9} (1999) 601
[arXiv:hep-ph/9811349].

\bibitem{Shoshi:2002in}
A.~I.~Shoshi, F.~D.~Steffen and H.~J.~Pirner,
Nucl.\ Phys.\ A {\bf 709} (2002) 131
[arXiv:hep-ph/0202012].

\bibitem{Gotsman:1999mw}
E.~Gotsman, E.~Levin, U.~Maor and E.~Naftali,
Eur.\ Phys.\ J.\ C {\bf 10} (1999) 689
[arXiv:hep-ph/9904277].

\bibitem{Cvetic:1999fi}
G.~Cvetic, D.~Schildknecht and A.~Shoshi,
Eur.\ Phys.\ J.\ C {\bf 13} (2000) 301
[arXiv:hep-ph/9908473].

\bibitem{McDermott:1999fa}
M.~McDermott, L.~Frankfurt, V.~Guzey and M.~Strikman,
Eur.\ Phys.\ J.\ C {\bf 16} (2000) 641
[arXiv:hep-ph/9912547].

\bibitem{Iancu:2003ge}
E.~Iancu, K.~Itakura and S.~Munier,
Phys.\ Lett.\ B {\bf 590} (2004) 199
[arXiv:hep-ph/0310338].

\bibitem{Kowalski:2006hc}
H.~Kowalski, L.~Motyka and G.~Watt,
Phys.\ Rev.\ D {\bf 74} (2006) 074016
[arXiv:hep-ph/0606272].

\bibitem{Golec-Biernat:2006ba}
K.~Golec-Biernat and S.~Sapeta,
Phys.\ Rev.\ D {\bf 74} (2006) 054032
[arXiv:hep-ph/0607276].

\bibitem{Bartels:2002cj}
J.~Bartels, K.~Golec-Biernat and H.~Kowalski,
Phys.\ Rev.\ D {\bf 66} (2002) 014001
[arXiv:hep-ph/0203258].

\bibitem{Bartels:2003yj}
J.~Bartels, K.~Golec-Biernat and K.~Peters,
Acta Phys.\ Polon.\ B {\bf 34} (2003) 3051
[arXiv:hep-ph/0301192].

\bibitem{Kowalski:2003hm}
H.~Kowalski and D.~Teaney,
Phys.\ Rev.\ D {\bf 68} (2003) 114005
[arXiv:hep-ph/0304189].

\bibitem{Ewerz:2006vd}
C.~Ewerz and O.~Nachtmann,
arXiv:hep-ph/0604087.

\bibitem{Nachtmann:ua}
O.~Nachtmann,
Annals Phys.\  {\bf 209} (1991) 436.

\bibitem{Nachtmann:1996kt}
O.~Nachtmann, 
Lectures given at 35th Internationale Universit\"atswochen 
f\"ur Kern- und Teilchenphysik, Schladming 1996, 
in {\sl Perturbative and Nonperturbative Aspects of Quantum Field Theory}, 
eds.\ H.\ Latal and W.\ Schweiger, Springer Verlag, Berlin, Heidelberg, 1997
[arXiv:hep-ph/9609365].

\bibitem{Hebecker:1997rv}
A.~Hebecker and P.~V.~Landshoff,
Phys.\ Lett.\ B {\bf 419} (1998) 393
[arXiv:hep-ph/9710296].

\bibitem{Buchmuller:1996xw}
W.~Buchm\"uller, M.~F.~McDermott and A.~Hebecker,
Nucl.\ Phys.\ B {\bf 487} (1997) 283
[Erratum-ibid.\ B {\bf 500} (1997) 621]
[arXiv:hep-ph/9607290].

\bibitem{Hebecker:1999ej}
A.~Hebecker,
Phys.\ Rept.\  {\bf 331} (2000) 1
[arXiv:hep-ph/9905226].

\bibitem{McLerran:1998nk}
L.~D.~McLerran and R.~Venugopalan,
Phys.\ Rev.\ D {\bf 59} (1999) 094002
[arXiv:hep-ph/9809427].

\bibitem{Venugopalan:1999wu}
R.~Venugopalan,
Acta Phys.\ Polon.\ B {\bf 30} (1999) 3731
[arXiv:hep-ph/9911371].

\bibitem{4}
O.\ Nachtmann, {\sl Elementary Particle Physics: Conpects and Phenomena}, 
Springer Verlag, Berlin, Heidelberg, 1990.

\bibitem{Chung:wm}
Y.~Chung, H.~G.~Dosch, M.~Kremer and D.~Schall,
Phys.\ Lett.\ B {\bf 102} (1981) 175.

\bibitem{Chung:cc}
Y.~Chung, H.~G.~Dosch, M.~Kremer and D.~Schall,
Nucl.\ Phys.\ B {\bf 197} (1982) 55.

\bibitem{Ioffe:kw}
B.~L.~Ioffe,
Nucl.\ Phys.\ B {\bf 188} (1981) 317
[Erratum-ibid.\ B {\bf 191} (1981) 591].

\bibitem{Lehmann:1954rq}
H.~Lehmann, K.~Symanzik and W.~Zimmermann,
Nuovo Cim.\  {\bf 1} (1955) 205.

\bibitem{Kuraev:fs}
E.~A.~Kuraev, L.~N.~Lipatov and V.~S.~Fadin,
Sov.\ Phys.\ JETP {\bf 45} (1977) 199
[Zh.\ Eksp.\ Teor.\ Fiz.\  {\bf 72} (1977) 377].

\bibitem{Balitsky:ic}
I.~I.~Balitsky and L.~N.~Lipatov,
Sov.\ J.\ Nucl.\ Phys.\  {\bf 28} (1978) 822
[Yad.\ Fiz.\  {\bf 28} (1978) 1597].

\bibitem{5}
J.\,D.\ Bjorken and S.\,D.\ Drell, 
{\sl Relativistic Quantum Fields}, Mc Graw Hill, 1965.

\bibitem{Nachtmann:2000gm}
O.~Nachtmann and A.~Rauscher,
Eur.\ Phys.\ J.\ C {\bf 16} (2000) 665
[arXiv:hep-ph/0002251].

\bibitem{Lepage:1979zb}
G.~P.~Lepage and S.~J.~Brodsky,
Phys.\ Lett.\ B {\bf 87} (1979) 359.

\bibitem{Lepage:1980fj}
G.~P.~Lepage and S.~J.~Brodsky,
Phys.\ Rev.\ D {\bf 22} (1980) 2157.

\end{thebibliography}
\end{document}